\documentclass[prd,aps,nofootinbib,preprintnumbers,amsmath,amssymb,floatfix]{revtex4}

\usepackage[a0paper]{geometry}
\usepackage{anysize}
\marginsize{2.5cm}{2cm}{2.5cm}{2.5 cm}
\usepackage{amsmath}
\usepackage{amsfonts}
\usepackage{amssymb}
\usepackage{slashed}
\usepackage{mathrsfs}
\usepackage{epsfig}
\usepackage{enumerate}
\usepackage{amsmath}
\usepackage{pdfpages}
\usepackage{setspace}
\usepackage[utf8]{inputenc}
\usepackage{float}
\usepackage{ytableau}

\newcommand{\pr}[1]{\mathcal{P}^{(#1)}}
\newcommand{\prt}[1]{\tilde{\mathcal{P}}^{(#1)}}

\usepackage{times}
\usepackage{hyperref}

\usepackage{graphicx}
\usepackage{dcolumn}
\usepackage{bm}
\usepackage{slashed}

\begin{document}

\title{Spin and flavor projection operators in the $SU(2N_f)$ spin-flavor group}

\author{
V{\'\i}ctor Miguel Banda Guzm\'an
}
\affiliation{
Instituto de F{\'i}sica y Matem\'aticas, Universidad Michoacana de San Nicol\'as de Hidalgo, Edificio C-3, Apdo.\ Postal 2-82, Morelia, Michoac\'an, 58040, M\'exico
}

\author{
Rub\'en Flores-Mendieta
}
\affiliation{
Instituto de F{\'\i}sica, Universidad Aut\'onoma de San Luis Potos{\'\i}, \'Alvaro Obreg\'on 64, Zona Centro, San Luis Potos{\'\i}, S.L.P.\ 78000, M\'exico
}

\author{
Johann Hern\'andez
}
\affiliation{
Instituto de F{\'\i}sica, Universidad Aut\'onoma de San Luis Potos{\'\i}, \'Alvaro Obreg\'on 64, Zona Centro, San Luis Potos{\'\i}, S.L.P.\ 78000, M\'exico
}

\author{
Felipe de Jes\'us Rosales-Aldape
}
\affiliation{
Instituto de F{\'\i}sica, Universidad Aut\'onoma de San Luis Potos{\'\i}, \'Alvaro Obreg\'on 64, Zona Centro, San Luis Potos{\'\i}, S.L.P.\ 78000, M\'exico
}

\date{\today}

\begin{abstract}

The quadratic Casimir operator of the special unitary $SU(N)$ group is used to construct projection operators, which can decompose any of its reducible finite-dimensional representation spaces contained in the tensor product of two and three adjoint spaces into irreducible components. Although the method is general enough, it is specialized to the $SU(2N_f) \to SU(2)\otimes SU(N_f)$ spin-flavor symmetry group, which emerges in the baryon sector of QCD in the large-$N_c$ limit, where $N_f$ and $N_c$ are the numbers of light quark flavors and color charges, respectively. The approach leads to the construction of spin and flavor projection operators that can be implemented in the analysis of the $1/N_c$ operator expansion. The use of projection operators allows one to successfully project out the desired components of a given operator and subtract off those that are not needed. Some explicit examples in $SU(2)$ and $SU(3)$ are detailed.

\end{abstract}

\maketitle

\section{Introduction}

The concept of symmetry, and specially gauge symmetry, is crucial in elementary particle physics. Early analyses of atomic spectra successfully implemented the use of $SU(2)$ representation theory to study the spin of particles. Further analyses in nuclear physics struggled to find out how protons and neutrons interact via a strong force to bind together into nuclei. Promptly it was discovered that the strong force had an $SU(2)$ invariance; it was called isospin symmetry and its irreducible representations (irreps) were labeled by isospin $1/2,1,\ldots$ A well-known example is the two-dimensional isospin-$1/2$ representation made up by the proton and neutron.

In the early decade of the 60s of the past century, a large number of new strongly interacting particles were discovered so it was imperative to classify them. Gell-Mann first suggested that they could be accommodated into irreps of $SU(3)$, so he proposed an organizational scheme for hadrons. It was called the \textit{eightfold way} \cite{gell-mann}; this peculiar name, presumably, is closely related to the fact that Gell-Mann mainly used the eight-dimensional adjoint representation of $SU(3)$.

Eventually, it was evident that the $SU(3)$ symmetry found by Gell-Mann was due to the existence of the three light quarks, $u$, $d$, $s$, which fitted into the fundamental three-dimensional representation of $SU(3)$. This symmetry has been since referred to as $SU(3)$ flavor symmetry. Hadrons were thus organized into $SU(3)$ representation multiplets---octets and decuplets---of roughly the same mass.

The special unitary group also plays a role in the local $SU(3)\otimes SU(2) \otimes U(1)$ gauge symmetry, which defines the modern standard model (SM) of particles and their interactions. Roughly, the three factors of the gauge symmetry give rise to the three fundamental interactions. Quantum chromodynamics (QCD), the theory of the strong interactions, is the $SU(3)$ component of the SM. It is a gauge theory of fermions---the quarks---and gauge bosons---the gluons---and stems from the fact that each quark comes in three completely identical states called colors; the symmetry is thus referred to as $SU(3)$ color symmetry. Unlike flavor symmetry, which is an approximate symmetry due to the relatively small masses of the three light quarks and plays a marginal role in the SM, color symmetry is exact and does play a preponderant role. At low energies, the running coupling constant of the theory is large, and the colored quarks and gluons must clump together to form colorless hadrons.

Various attempts have been made so far to construct grand unified theories of the weak, strong, and electromagnetic interactions. These approaches mostly use Lie groups. Common examples are $SU(5)$ in the simplest grand unification theory, $SO(10)$, and $E6$. Further applications of $SU(N)$ can also be found in shell models of nuclear and atomic physics \cite{Cseh,shell_at}, the worldline approach to non-Abelian gauge fields \cite{Corradini_1, James_1, James_2}, to name but a few.

It should be stressed that, despite the tremendous progress achieved in the understanding of the strong interactions with QCD, the analytical calculation of the structure and interactions of hadrons directly in terms of the underlying quark-gluon dynamics is not possible because the theory is strongly coupled at low energies. Soon after the advent of QCD, ’t Hooft pointed out that gauge theories based on the $SU(N_c)$ group simplify in the limit $N_c \to \infty$, where $N_c$ is the number of color charges \cite{thooft1}. Baryons in large-$N_c$ QCD were first studied by Witten \cite{witten}. Later, it was shown that in the large-$N_c$ limit the baryon sector has an exact contracted $SU(2N_f)$ spin-flavor symmetry, where $N_f$ is the number of light quark flavors \cite{dm1,dm2,gs1,gs2}. Physical quantities are then considered in this limit, where corrections emerge at relative orders $1/N_c$, $1/N_c^2$, and so on; this sequence originates the $1/N_c$ expansion of QCD.

The $1/N_c$ expansion turns out to be quite useful for studying the interactions and properties of large-$N_c$ color-singlet baryons at low energies. The construction of the $1/N_c$ expansion of any QCD operator transforming according to a given spin$\otimes$flavor representation is expressed in terms of $n$-body operators $O_n$, which can be written as polynomials of homogeneous degree $n$ in the spin-flavor generators. The operators $O_n$ make up a complete and independent operator basis \cite{djm95}. It should be emphasized that for baryons at large finite $N_c$, the $1/N_c$ operator expansion only extends to $N_c$-body operators in the baryon spin-flavor generators. Although straightforward in principle, the reduction of higher-order operator structures to the physical operator basis turns out to be quite tedious due to the considerable amount of group theory involved. The fact that the operator basis is complete and independent makes those reductions possible.

Here is precisely where the aim of the present paper can be delineated: to present a general procedure to construct projection operators in $SU(N)$ out of the corresponding Casimir operators. The projection operators so obtained act on tensor operators that belong to tensor products of adjoint representation spaces, decomposing them into different operators with specific quadratic Casimir eigenvalues. The applicability to the $1/N_c$ operator expansion is immediate. The cases of physical interest for $N_f=2$ and $N_f=3$ are worked out to show the usefulness of the resultant projectors. In passing, it can be pointed out that the method is not limited to the $1/N_c$ expansion, but it can also be used in shell models of atomic and nuclear physics; in this case, the projector method allows one to construct tensor operators which, with the aid of the Wigner-Eckart theorem, can be used to calculate transition amplitudes. The worldline approach to non-Abelian gauge fields is also another area where the projector method can be adapted to fit there. All in all, the method shows some potential applicability in areas where the $SU(N)$ group is involved.

The organization of the paper is as follows. In Sec.~\ref{sec:general}, some theoretical aspects of the $SU(N)$ group are briefly summarized, starting with some rather elementary concepts and definitions, which are provided to set notation and conventions. A key feature in the analysis is the definition of the adjoint space and the tensor space formed by the product of $n$ adjoint spaces. The latter can always be decomposed into subspaces labeled by a specific eigenvalue of the quadratic Casimir operator of the algebra of $SU(N)$. The procedure to do so is discussed at the end of this section, and the defining general expression of the projection operator is provided. In Sec.~\ref{sec:2adj}, the projection operators for the tensor product space of two adjoint spaces are constructed explicitly. The properties that by definition projection operators are demanded to fulfilled are rigorously verified. The particular case $N=2$ is also discussed at the end of this section. In Sec.~\ref{sec:2asu3}, the projection operators previously defined are specialized to the $SU(2N_f)$ spin-flavor symmetry group, which breaks to its spin and flavor groups $SU(2)\otimes SU(N_f)$. Consequently, the spin and flavor projection operators are constructed and readily applied to the $1/N_c$ operator expansion. In Sec.~\ref{sec:3adj}, the method is outlined for the tensor product space of three adjoint spaces. In this case, the explicit construction of projection operators becomes a rather involved task, so only a few examples are detailed. Some closing remarks and conclusions are provided in Sec.~\ref{sec:clo}. The paper is complemented by two appendices, where some supplemental information is provided.

\section{\label{sec:general}Projector technique for $SU(N)$ adjoint tensor operators}

To start with, a salient definition is that of a Lie group. It is defined as a group in which the elements are labeled by a set of continuous parameters with a multiplication law that depends smoothly on the parameters themselves \cite{georgi}. A compact Lie group, on the other hand, is a Lie group in which the parametrization consists of a finite number of bounded parameter domains; otherwise, the group is referred to as noncompact \cite{greiner}. The $SU(N)$ group of all complex unitary matrices of order $N$ with determinant 1 and the $SO(N)$ group of all real orthogonal matrices of order $N$ with determinant 1 are two well-known examples of connected compact Lie groups.

The elements of a Lie group can be written as
\begin{equation}
\exp{\left[i\sum_a \beta^a X^a \right]},
\end{equation}
where $\beta^a$, $a=1,\ldots,N$ are real numbers and $X^a$ are linearly independent Hermitian operators. Hereafter, and unless explicitly noticed otherwise, the sum over repeated indices will be implicit. The $X^a$ are referred to as the generators of the Lie group and they satisfy the commutation relations
\begin{equation}
[X^a,X^b] = i f^{abc} X^c. \label{eq:comm}
\end{equation}
The $f^{abc}$ are referred to as the structure constants of the Lie group. The vector space $\beta^a X^a$, together with the commutation relations (\ref{eq:comm}), define the Lie algebra associated with the Lie group.

The generators satisfy the Jacobi identity,
\begin{equation}
[X^a,[X^b,X^c]] + \mbox{cyclic permutations} = 0,
\end{equation}
which in terms of the structure constants becomes
\begin{equation}
f^{bce}f^{aeg} + f^{abe}f^{ceg} + f^{cae}f^{beg} = 0.
\end{equation}

The quadratic Casimir operator is defined as
\begin{equation}
C \equiv X^e X^e,
\end{equation}
so that
\begin{equation}
[C,X^a] = 0.
\end{equation}

As for the $SU(N)$ group, let $T^a$ be operators that generate the Lie algebra of the group. There are $N^2-1$ of such operators, which serve as a basis for the set of traceless Hermitian $N\times N$ matrices. The generators satisfy the commutation relations
\begin{equation}
[T^a,T^b] = i f^{abc} T^c,
\end{equation}
where $a,b,c$ run from 1 to the dimension of the Lie algebra of $SU(N)$, i.e., from 1 to $N^2-1$.

In the fundamental representation of $SU(N)$, the normalization convention usually adopted for the generators reads
\begin{equation}
\mathrm{Tr}(T^a T^b) = \frac12 \delta^{ab},
\end{equation}
so in this convention the $f^{abc}$ are totally antisymmetric with respect to the interchange of any two indices.

Let $T_A^a$ define a set of operators such that
\begin{equation}
\left[ T_A^a \right]^{cb} \equiv i f^{abc}, \label{eq:adjoin_rep}
\end{equation}
i.e., the structure constants themselves constitute a matrix representation of the operators. The representation generated by the structure constants is called the \textit{adjoint representation}.

An $SU(N)$ adjoint operator $Q^a$ can thus be defined, such that
\begin{equation}
[T^a,Q^b] = i f^{abc} Q^c. \label{eq:sun_op}
\end{equation}
The operators $Q^a$ can make up a basis for the carrier space where the generators of the Lie algebra of $SU(N)$ in the adjoint representation act \cite{greiner}. If $T_A^a$ are taken as the generators in the adjoint representation, relation (\ref{eq:sun_op}) is equivalent to
\begin{equation}
T_A^a Q^b = i f^{abc} Q^c. \label{eq:T_AonQ}
\end{equation}
Hereafter, the carrier space generated by the operators $Q^a$ will be referred to as the \textit{adjoint space} and will be denoted by $adj = \{Q^a\}$.

Another tensor space of interest is the one formed by the product of the adjoint space with itself $n$ times. It is denoted by $\prod_{i=1}^n adj \otimes$. This space can usually be decomposed into subspaces labeled by a specific eigenvalue of the quadratic Casimir operator $C$ of the Lie algebra of $SU(N)$. The decomposition can be achieved by adapting the projector technique for decomposing reducible representations introduced in Ref.~\cite{Guzman:2019}. Following the lines of that reference, the sought projection operators $\pr{m}$ are thus constructed as
\begin{equation}
\pr{m} = \prod_{i=1}^k \left[ \frac{C - c_{n_i}}{c_{m} - c_{n_i}} \right], \qquad \qquad c_m \neq c_{n_i}, \label{eq:general_proj}
\end{equation}
where $k$ labels the number of different possible eigenvalues for the quadratic Casimir operator and $c_{m}$ are its eigenvalues given by \cite{Gross},
\begin{equation}
c_m = \frac12 \left[n N - \frac{n^2}{N} + \sum_i r_i^2 - \sum_i c^2_i \right], \label{eq:egenvalues_C}
\end{equation}
where $n$ is the total number of boxes of the Young tableu for a specific representation, $r_i$ is the number of boxes in the $i$th row and $c_i$ is the number of boxes in the $i$th column.

From the defining expression (\ref{eq:general_proj}), it can be inferred that if $\prod_{i=1}^n Q_i^{a_i}$ is an $SU(N)$ tensor operator, where each $Q_i^{a_i}$ satisfies the commutation relation (\ref{eq:sun_op}), then
\begin{equation}
\pr{m} \prod_{i=1}^n Q_i^{a_i} = \tilde{Q}^{a_1 \dots a_n},
\end{equation}
where the tensor $\tilde{Q}^{a_1 \dots a_n}$ is an eigenstate for the quadratic Casimir $C$ with eigenvalue $c_m$,
\begin{equation}
C \, \tilde{Q}^{a_1 \dots a_n} = c_m \, \tilde{Q}^{a_1 \dots a_n}.
\end{equation}

In the following sections, the decompositions of the tensor spaces $adj \otimes adj$ and $adj \otimes adj \otimes adj$ will be carried out by using the projector technique described above.

\section{\label{sec:2adj}Projection operators in the tensor space $adj \otimes adj$}

The Young tableau for the adjoint representation is given by
\begin{equation*}
adj \; = \;\; \ytableausetup{mathmode, boxsize=1.8em}
\begin{ytableau}
\scriptstyle 1 & \scriptstyle 2 \\
\scriptstyle 2 \\
\none[\vdots] \\
\scriptstyle N-1
\end{ytableau}
\end{equation*}

The tensor space $adj \otimes adj$ decomposes as
\begin{eqnarray}
adj \otimes adj = 1 \; \oplus \; 2 \quad
\ytableausetup{mathmode, boxsize=1.8em}
\begin{ytableau}
\scriptstyle 1 & \scriptstyle 2 \\
\scriptstyle 2 \\
\scriptstyle 3 \\
\none[\vdots] \\
\scriptstyle N-1
\end{ytableau}
\; \oplus \;
\ytableausetup{mathmode, boxsize=1.8em}
\begin{ytableau}
\scriptstyle 1 & \scriptstyle 2 & \scriptstyle 3 \\
\scriptstyle 2 \\
\none[\vdots] \\
\scriptstyle N-2
\end{ytableau}
\; \oplus \;
\ytableausetup{mathmode, boxsize=1.8em}
\begin{ytableau}
\scriptstyle 1 & \scriptstyle 2 & \scriptstyle 3 \\
\scriptstyle 2 & \scriptstyle 3 & \scriptstyle 4 \\
\scriptstyle 3 & \scriptstyle 4 \\
\none[\vdots] \\
\scriptstyle N-1 & \scriptstyle N
\end{ytableau}
\; \oplus \;
\ytableausetup{mathmode, boxsize=1.8em}
\begin{ytableau}
\scriptstyle 1 & \scriptstyle 2 & \scriptstyle 3 & \scriptstyle 4 \\
\scriptstyle 2 & \scriptstyle 3 \\
\scriptstyle 3 & \scriptstyle 4 \\
\none[\vdots] \\
\scriptstyle N-1 & \scriptstyle N
\end{ytableau}
\; \oplus \;
\ytableausetup{mathmode, boxsize=1.8em}
\begin{ytableau}
\scriptstyle 1 & \scriptstyle 2 \\
\scriptstyle 2 & \scriptstyle 3 \\
\scriptstyle 3 \\
\none[\vdots] \\
\scriptstyle N-2
\end{ytableau} \vspace{15pt} \nonumber
\end{eqnarray}

In the notation of Ref.~\cite{djm95}, the above irreps are designated by
\begin{equation}
adj \otimes adj = 1 \oplus 2 adj \oplus \bar{a}s \oplus \bar{s}a \oplus \bar{s}s \oplus \bar{a}a, \label{eq:adjd}
\end{equation}
so this convenient notation will also be used here.

The quadratic Casimir eigenvalues for each representation in the decomposition of $adj \otimes adj$ are obtained from Eq.~(\ref{eq:egenvalues_C}) and are listed in the second column (from left to right) of Table \ref{tab:1}.

\begin{table}[h!]
\centering
\begin{tabular}{|c|c|c|}
\hline
Rep & Eigenvalue & Representation \\
\hline &  & \\
1 & $c_0 = 0$ & 1 \\
&  & \\ \hline
&  & \\
$adj$ & $c_1 = N$ &
\footnotesize \ytableausetup{mathmode, boxsize=2.1em}
\begin{ytableau}
1 & 2 \\
2 \\
3 \\
\none[\vdots] \\
\scriptstyle N-1
\end{ytableau}
\\
&  & \\ \hline
&  & \\
\quad $\bar{a}s \oplus \bar{s}a \quad $ & $c_2 = 2N$ & \quad
\footnotesize
\ytableausetup{mathmode, boxsize=2.2em}
\begin{ytableau}
1 & 2 & 3 \\
2 \\
\none[\vdots] \\
\scriptstyle N-2
\end{ytableau}
\,\,\, $\oplus$ \,\,
\ytableausetup{mathmode, boxsize=2.2em}
\begin{ytableau}
1 & 2 & 3 \\
2 & 3 & 4 \\
3 & 4 \\
\none[\vdots] \\
\scriptstyle N-1 & \scriptstyle N
\end{ytableau}
\quad \qquad \\
& & \\ \hline
& & \\
$\bar{s}s$ & \quad $c_3 = 2(N+1)$ \quad \quad &
\footnotesize \ytableausetup{mathmode, boxsize=2.2em}
\begin{ytableau}
1 & 2 & 3 & 4 \\
2 & 3 \\
3 & 4 \\
\none[\vdots] \\
\scriptstyle N-1 & \scriptstyle N
\end{ytableau}
\\
& & \\ \hline
& & \\
$\bar{a}a$ & $c_4 = 2(N-1)$ &
\footnotesize \ytableausetup{mathmode, boxsize=2.2em}
\begin{ytableau}
1 & 2 \\
2 & 3 \\
3 \\
\none[\vdots] \\
\scriptstyle N-2
\end{ytableau}
\\
& & \\ \hline
\end{tabular}
\caption{\label{tab:1}Quadratic Casimir eigenvalues and projectors corresponding to each representation in the decomposition of the reducible tensor representation $adj \otimes adj$.}
\end{table}

Since five different eigenvalues are available, the projectors in Eq.~(\ref{eq:general_proj}) are computed as
\begin{equation}
\pr{m} = \frac{ \alpha_0 - \alpha_1 C + \alpha_2 C^2 - \alpha_3 C^3 + C^4 }{\prod_{i=1}^4 (c_m - c_{n_i})}, \qquad \qquad c_m \neq c_{n_i}, 
\label{eq:prmfin}
\end{equation}
where
\begin{subequations}
\begin{equation}
\alpha_0 = c_{n_1}c_{n_2}c_{n_3}c_{n_4},
\end{equation}
\begin{equation}
\alpha_1 = c_{n_1}c_{n_2}c_{n_3} + c_{n_1}c_{n_2}c_{n_4} + c_{n_1}c_{n_3}c_{n_4} + c_{n_2}c_{n_3}c_{n_4},
\end{equation}
\begin{equation}
\alpha_2 = c_{n_1}c_{n_2} + c_{n_1}c_{n_3} + c_{n_1}c_{n_4} + c_{n_2}c_{n_3} + c_{n_2}c_{n_4} + c_{n_3}c_{n_4},
\end{equation}
and
\begin{equation}
\alpha_3 = c_{n_1} + c_{n_2} + c_{n_3} + c_{n_4}.
\end{equation}
\end{subequations}

A word of caution is in order here. The defining expression of $\pr{m}$, Eq.~(\ref{eq:general_proj}), and its subsequent version for the tensor space $adj\otimes adj$, Eq.~(\ref{eq:prmfin}), impose the condition $c_m \neq c_{n_i}$ in order to avoid singularities. Particularly, note that $\bar{a}s$ and $\bar{s}a$ are complex-conjugated representations, so they share the same eigenvalue of the Casimir operator, $c_2=2N$, according to Table \ref{tab:1}. For this reason, it is not only convenient but also necessary to construct a projection operator that comprises both representations, as it is described below.

On the other hand, a complete determination of $C$ demands the evaluation of the generators $T_{2A}^a$ that act in the tensor space $adj \otimes adj$. In terms of $T_A$ they are given by
\begin{equation}
T_{2A}^a = T_A^a \otimes \openone + \openone \otimes T_A^a. \label{eq:T2A_t}
\end{equation}
Therefore, 
\begin{eqnarray}
C & = & T_{2A}^e T_{2A}^e \nonumber \\
& = & T_A^e T_A^e \otimes \openone + \openone \otimes T_A^e T_A^e + 2 T_A^e \otimes T_A^e.
\end{eqnarray}

Since $T_A^e T_A^e$ is the quadratic Casimir operator for the adjoint representation, then by Schur's lemma,
\begin{equation}
T_A^e T_A^e = N \openone. \label{eq:casimir_1}
\end{equation}
Thus,
\begin{equation}
C = 2(N \openone \otimes \openone + T_A^e \otimes T_A^e). \label{eq:Ct}
\end{equation}

The action of $C$ on a tensor operator $Q_{1}^{b_1}Q_{2}^{b_2}$ yields, according to Eq.~(\ref{eq:T_AonQ}),
\begin{equation}
C Q_{1}^{b_1}Q_{2}^{b_2} = 2 (N \delta^{b_1a_1} \delta^{b_2a_2} - f^{b_1a_1e} f^{b_2a_2e}) Q_{1}^{a_1}Q_{2}^{a_2}. \label{eq:C_QQ}
\end{equation}

Therefore, in components, $C$ reads
\begin{eqnarray}
\left[ C \right]^{a_1a_2b_1b_2} & = & 2 (N \delta^{b_1a_1} \delta^{b_2a_2} - F^{b_1a_1b_2a_2}) \nonumber \\
& = & \frac{4}{N} \left[ \frac{N^2}{2} \delta^{b_1a_1} \delta^{b_2a_2} - \delta^{b_1b_2} \delta^{a_1a_2} + \delta^{b_1a_2} \delta^{b_2a_1} \right] - 2 (D^{b_1b_2a_1a_2} - D^{b_1a_2b_2a_1}), \label{eq:C_c}
\end{eqnarray}
where the second equality follows from the identities listed in Appendix \ref{sec:idenSUN}. For the ease of notation, the symbols $F^{a_1a_2b_1b_2}$ and $D^{a_1a_2b_1b_2}$ have also been introduced; they read
\begin{subequations}
\begin{equation}
F^{a_1a_2b_1b_2} = f^{a_1a_2e} f^{b_1b_2e}, \label{eq:Def_F}
\end{equation}
\begin{equation}
D^{a_1a_2b_1b_2} = d^{a_1a_2e} d^{b_1b_2e}, \label{eq:Def_D}
\end{equation}
\end{subequations}
where the fully symmetric coefficients $d^{a_1a_2a_3}$ read
\begin{equation}
d^{a_1a_2a_3} = \frac14 \mathrm{Tr}\, (\{T^{a_1},T^{a_2}\}T^{a_3}).
\end{equation}

Additionally, let us also define the operator $G$ acting on $Q_{1}^{b_1}Q_{2}^{b_2}$ with components,
\begin{equation}
[G]^{a_1a_2b_1b_2} = - F^{a_1b_1a_2b_2}, \label{eq:def_T}
\end{equation}
so that the quadratic Casimir $C$ in Eq.~(\ref{eq:Ct}) can be rewritten as
\begin{equation}
C = 2(N + G), \label{eq:CC}
\end{equation}
and powers of $C$ are straightforwardly obtained as
\begin{equation}
C^2 = 4 (N^2 + 2 N G + G^2), \label{eq:C2}
\end{equation}
\begin{equation}
C^3 = 8 (N^3 + 3 N^2 G + 3 N G^2 + G^3), \label{eq:C3}
\end{equation}
\begin{equation}
C^4 = 16 (N^4 + 4 N^3 G + 6 N^2 G^2 + 4 N G^3 + G^4). \label{eq:C4}
\end{equation}

By making use again of the identities listed in Appendix \ref{sec:idenSUN}, the powers of the operator $G$ required in the analysis are explicitly given by
\begin{subequations}
\begin{equation}
\left[ G^2 \right]^{a_1a_2b_1b_2} = \frac12 (2 \delta^{a_1a_2} \delta^{b_1b_2} + \delta^{a_1b_1} \delta^{a_2b_2} + \delta^{a_1b_2} \delta^{a_2b_1}) + \frac{N}{4} (F^{a_1a_2b_1b_2} + D^{a_1a_2b_1b_2}),
\end{equation}
\begin{equation}
\left[ G^3 \right]^{a_1a_2b_1b_2} = - N \delta^{a_1a_2} \delta^{b_1b_2} - \frac12 (F^{a_1b_1a_2b_2} + F^{a_2b_1a_1b_2}) - \frac{N^2}{8} (F^{a_1a_2b_1b_2} + D^{a_1a_2b_1b_2}),
\end{equation}
and
\begin{equation}
\left[ G^4 \right]^{a_1a_2b_1b_2} = (N^2+1) \delta^{a_1a_2} \delta^{b_1b_2} + \frac12 (\delta^{a_1b_1} \delta^{a_2b_2} + \delta^{a_2b_1} \delta^{a_1b_2}) + \frac{N^3}{16} F^{a_1a_2b_1b_2} + \frac{1}{16}N(N^2+4) D^{a_1a_2b_1b_2}.
\end{equation}
\end{subequations}

All the necessary powers of $C$ involved in Eq.~(\ref{eq:general_proj}) are now explicitly determined so the projection operator $\pr{m}$, corresponding to eigenvalue $c_m$ of $C$, can be evaluated. For instance, for $c_0 = 0$,
\begin{eqnarray}
\pr{0} & = & \frac{\alpha_0 - \alpha_1 C + \alpha_2 C^2 - \alpha_3 C^3 + C^4 }{\prod_{i=1}^4(0 - c_i)}
\nonumber \\
& = & \mbox{} - \frac{N G + 2 G^2 - N G^3 - 2 G^4}{N^2(N^2-1)},
\end{eqnarray}
where
\begin{subequations}
\begin{equation}
\alpha_0 = 8 N^2(N^2-1),
\end{equation}
\begin{equation}
\alpha_1 = 4 N(5N^2-3),
\end{equation}
\begin{equation}
\alpha_2 = 18 N^2-4,
\end{equation}
and
\begin{equation}
\alpha_3 = 7 N.
\end{equation}
\end{subequations}

Thus,
\begin{equation}
\left[\pr{0}\right]^{a_1a_2b_1b_2} = \frac{1}{N^2-1} \delta^{a_1a_2} \delta^{b_1b_2}. \label{eq:p0}
\end{equation}

The procedure can be repeated for the remaining four eigenvalues, which yields the projection operators
\begin{equation}
\left[ \pr{1} \right]^{a_1a_2b_1b_2} = \frac{N}{N^2-4} D^{a_1a_2b_1b_2} + \frac{1}{N} F^{a_1a_2b_1b_2}, \label{eq:p1}
\end{equation}
\begin{equation}
\left[ \pr{2} \right]^{a_1a_2b_1b_2} = \frac12 (\delta^{a_1b_1} \delta^{a_2b_2} - \delta^{a_2b_1} \delta^{a_1b_2}) - \frac{1}{N} F^{a_1a_2b_1b_2}, \label{eq:p2}
\end{equation}
\begin{eqnarray}
\left[ \pr{3} \right]^{a_1a_2b_1b_2} & = & \frac{N+2}{4N} (\delta^{a_1b_1} \delta^{a_2b_2} + \delta^{a_2b_1} \delta^{a_1b_2}) - \frac{N+2}{2N(N+1)} \delta^{a_1a_2} \delta^{b_1b_2} - \frac{N+4}{4(N+2)} D^{a_1a_2b_1b_2} \nonumber \\
&  & \mbox{} + \frac14 (D^{a_1b_1a_2b_2} + D^{a_2b_1a_1b_2}), \label{eq:p3}
\end{eqnarray}
\begin{eqnarray}
\left[ \pr{4} \right]^{a_1a_2b_1b_2} & = & \frac{N-2}{4N} (\delta^{a_1b_1} \delta^{a_2b_2} + \delta^{a_2b_1} \delta^{a_1b_2}) + \frac{N-2}{2N(N-1)} \delta^{a_1a_2} \delta^{b_1b_2} + \frac{N-4}{4(N-2)} D^{a_1a_2b_1b_2} \nonumber \\
&  & \mbox{} - \frac14 (D^{a_1b_1a_2b_2} + D^{a_2b_1a_1b_2}). \label{eq:p4}
\end{eqnarray}

The above projection operators satisfy the properties
\begin{equation}
\left[\pr{m}\right]^{a_1a_2d_1d_2} \left[\pr{n}\right]^{d_1d_2b_1b_2} = \left\{
\begin{array}{ll}
0, & \quad m \neq n \\[3mm]
\displaystyle \left[\pr{m}\right]^{a_1a_2b_1b_2}, & \quad m = n,
\end{array}
\right.
\end{equation}
which are demanded by definition.

Also, notice that
\begin{equation}
\sum_{m=0}^4 \left[ \pr{m} \right]^{a_1a_2b_1b_2} = \delta^{a_1b_1} \delta^{a_2b_2},
\end{equation}
so they constitute a complete set of operators.

Now, given two adjoints $Q_{1}^{b_1}$ and $Q_{2}^{b_2}$, the action of projectors $\pr{m}$ on the adjoint tensor operator $Q_{1}^{b_1}Q_{2}^{b_2}$ yields
\begin{eqnarray}
\left[ Q^{(0)} \right]^{b_1b_2} & = & \left[\pr{0} Q_{1} Q_{2}\right]^{b_1b_2} \nonumber \\
& = & \mbox{} \frac{1}{N^2-1} \delta^{b_1b_2} Q_{1}^e Q_{2}^e, \label{eq:pqq0}
\end{eqnarray}
\begin{eqnarray}
\left[ Q^{(1)} \right]^{b_1b_2} & = & \left[\pr{1} Q_{1} Q_{2}\right]^{b_1b_2} \nonumber \\
& = & \mbox{} \frac{N}{N^2-4} D^{b_1b_2a_1a_2} Q_{1}^{a_1} Q_{2}^{a_2} + \frac{1}{N} F^{b_1b_2a_1a_2} Q_{1}^{a_1} Q_{2}^{a_2}, \label{eq:pqq1}
\end{eqnarray}
\begin{eqnarray}
\left[ Q^{(2)} \right]^{b_1b_2} & = & \left[\pr{2} Q_{1} Q_{2}\right]^{b_1b_2} \nonumber \\
& = & \mbox{} \frac12 \left(Q_{1}^{b_1} Q_{2}^{b_2} - Q_{1}^{b_2} Q_{2}^{b_1} \right) - \frac{1}{N} F^{b_1b_2a_1a_2} Q_{1}^{a_1} Q_{2}^{a_2}. \label{eq:pqq2}
\end{eqnarray}
\begin{eqnarray}
\left[ Q^{(3)} \right]^{b_1b_2} & = & \left[\pr{3} Q_{1} Q_{2}\right]^{b_1b_2} \nonumber \\
& = & \frac{N+2}{4N} \left(Q_{1}^{b_1} Q_{2}^{b_2} + Q_{1}^{b_2} Q_{2}^{b_1} \right) - \frac{N+2}{2N(N+1)} \delta^{b_1b_2} Q_{1}^{e} Q_{2}^{e} - \frac{N+4}{4(N+2)} D^{a_1a_2b_1b_2} Q_{1}^{a_1} Q_{2}^{a_2} \nonumber \\
&  & \mbox{} + \frac14 \left(D^{b_1a_1b_2a_2} + D^{b_1a_2b_2a_1} \right) Q_{1}^{a_1} Q_{2}^{a_2} , \label{eq:pqq3}
\end{eqnarray}
\begin{eqnarray}
\left[ Q^{(4)} \right]^{b_1b_2} & = & \left[\pr{4} Q_{1} Q_{2}\right]^{b_1b_2} \nonumber \\
& = & \frac{N-2}{4N} \left(Q_{1}^{b_1} Q_{2}^{b_2} + Q_{1}^{b_2} Q_{2}^{b_1} \right) + \frac{N-2}{2N(N-1)} \delta^{b_1b_2} Q_{1}^{e} Q_{2}^{e} + \frac{N-4}{4(N-2)} D^{a_1a_2b_1b_2} Q_{1}^{a_1} Q_{2}^{a_2} \nonumber \\
&  & \mbox{} - \frac14 \left(D^{b_1a_1b_2a_2} + D^{b_1a_2b_2a_1} \right) Q_{1}^{a_1} Q_{2}^{a_2}. \label{eq:pqq4}
\end{eqnarray}

The operators on the left-hand sides in Eqs.~(\ref{eq:pqq0}--\ref{eq:pqq4}) are labeled by an index that indicates the space representation they belong to. Therefore, when projection operator $\pr{m}$ acts on the tensor product of two adjoints, it projects out precisely the component of the representation it belongs to. Two simple examples for $N=2$ and $N=3$ suffice to illustrate the usefulness of the projection operators so far constructed. These examples are worked out in the following sections.

\subsection{\label{sec:2asu2}Projection operators for $N=2$}

$SU(2)$ is the simplest non-Abelian Lie group. It appears in two scenarios in physics. One is as the spin double cover of the rotation $SO(3)$ group, and the other is as an internal symmetry relating types of particles. Explicit realizations of them are spin and isotopic spin symmetries. The generators $J^i$ and $I^a$ correspond to spin and isospin, respectively, and the corresponding conventional structure constants are $\epsilon^{ijk}$ ($i,j,k=1,2,3$) and $\epsilon^{abc}$ ($a,b,c=1,2,3$), which are totally antisymmetric.

To construct the projection operators for $N=2$, an important issue to be kept in mind is the fact that $SU(2)$ does not admit representations for the eigenvalues $c_2=2N$ and $c_4=2(N-1)$ of the quadratic Casimir operator in the space $adj \otimes adj$ listed in Table \ref{tab:1}, so the procedure to construct $\pr{m}$ must be adapted accordingly because, in particular, $\pr{1}$ of Eq.~(\ref{eq:p1}) as it stands is ill-defined for $N=2$. Therefore, the procedure must be repeated accounting for the eigenvalues $c_0$, $c_1$, and $c_3$ only.

While the projector $\pr{0}$ is easily obtained as
\begin{equation}
\left[ \pr{0} \right]^{a_1a_2b_1b_2} = \frac13 \delta^{a_1a_2} \delta^{b_1b_2}, \label{eq:p0s}
\end{equation}
$\pr{1}$ is constructed as
\begin{equation}
\pr{1} = \frac{C^2 - (c_0+c_3)C+c_0c_3}{(c_1-c_0)(c_1-c_2)}.
\end{equation}

From $C$ and $C^2$ given in Eqs.~(\ref{eq:CC}) and (\ref{eq:C2}) for $N=2$, it follows that
\begin{equation}
\pr{1} = \frac18 (6C - C^2),
\end{equation}
so that
\begin{equation}
\left[ \pr{1} \right]^{a_1a_2b_1b_2} = \frac12 (\delta^{a_1b_1} \delta^{a_2b_2} - \delta^{a_2b_1} \delta^{a_1b_2}). \label{eq:p1s}
\end{equation}

Similarly,
\begin{equation}
\left[ \pr{3} \right]^{a_1a_2b_1b_2} = \frac12 (\delta^{a_1b_1} \delta^{a_2b_2} + \delta^{a_2b_1} \delta^{a_1b_2}) - \frac13 \delta^{a_1a_2} \delta^{b_1b_2}. \label{eq:p3s}
\end{equation}

Now, given two adjoints $Q_{1}^{b_1}$ and $Q_{2}^{b_2}$ defined in spin space, for instance, the projectors $\pr{0}$, $\pr{1}$, and $\pr{3}$, given by Eqs.~(\ref{eq:p0s}), (\ref{eq:p1s}), and (\ref{eq:p3s}), acting on the adjoint tensor operator $Q_{1}^{b_1}Q_{2}^{b_2}$, project out the $J=0$, $J=1$, and $J=2$ spin components of that tensor product, respectively. Similar conclusions can be reached for isospin space, of course.

\section{\label{sec:2asu3}Projection operators in $SU(2N_f) \to SU(2) \otimes SU(N_f)$ spin-flavor symmetry}

In the introductory section it was pointed out that the baryon sector of QCD has a contracted $SU(2N_f)$ symmetry, where $N_f$ is the number of light quark flavors \cite{dm1,dm2,gs1,gs2}. Under the decomposition $SU(2N_f) \to SU(2) \otimes SU(N_f)$, the spin-flavor representation yields a tower of baryon flavor representations with spins $J=1/2,3/2,\ldots,N_c/2$ \cite{djm95,gs1}. The spin-flavor generators of $SU(2N_f)$ can be written as one-body quark operators acting on the $N_c$-quark baryon states, namely,
\begin{subequations}
\label{eq:su6gen}
\begin{eqnarray}
J^k & = & \sum_\alpha^{N_c} q_\alpha^\dagger \left(\frac{\sigma^k}{2}\otimes \openone \right) q_\alpha, \\
T^c & = & \sum_\alpha^{N_c} q_\alpha^\dagger \left(\openone \otimes \frac{\lambda^c}{2} \right) q_\alpha, \\
G^{kc} & = & \sum_\alpha^{N_c} q_\alpha^\dagger \left(\frac{\sigma^k}{2}\otimes \frac{\lambda^c}{2} \right) q_\alpha.
\end{eqnarray}
\end{subequations}
Here $q_\alpha^\dagger$ and $q_\alpha$ constitute a set of quark creation and annihilation operators, where $\alpha=1,\ldots,N_f$ denote the $N_f$ quark flavors with spin up and $\alpha=N_f+1,\ldots,2N_f$ the $N_f$ quark flavors with spin down. Likewise, $J^k$ are the spin generators, $T^c$ are the flavor generators, and $G^{kc}$ are the spin-flavor generators. The $SU(2N_f)$ spin-flavor generators satisfy the commutation relations listed in Table \ref{tab:surel} \cite{djm95}.
\begingroup
\begin{table}
\bigskip
\centerline{\vbox{ \tabskip=0pt \offinterlineskip
\halign{
\strut\quad $ # $\quad\hfil&\strut\quad $ # $\quad \hfil\cr
\multispan2\hfil $\left[J^i,T^a\right]=0,$ \hfil \cr
\noalign{\medskip}
\left[J^i,J^j\right] = i\epsilon^{ijk} J^k,
&\left[T^a,T^b\right] = i f^{abc} T^c,\cr
\noalign{\medskip}
\left[J^i,G^{ja}\right] = i\epsilon^{ijk} G^{ka},
&\left[T^a,G^{ib}\right] = i f^{abc} G^{ic},\cr
\noalign{\medskip}
\multispan2\hfil$\displaystyle [G^{ia},G^{jb}] = \frac{i}{4}\delta^{ij}
f^{abc} T^c + \frac{i}{2N_f} \delta^{ab} \epsilon^{ijk} J^k + \frac{i}{2} \epsilon^{ijk} d^{abc} G^{kc}.$ \hfill\cr }}}
\caption{\label{tab:surel}$SU(2N_f)$ commutation relations.}
\end{table}
\endgroup

The approach to obtain projection operators discussed in the previous sections can now be implemented to the $SU(2N_f)$ spin-flavor symmetry to construct spin and flavor projection operators, which will act on well-defined $n$-body operators. For the ease of notation, throughout this section lowercase letters ($i,j,\ldots$) will denote indices transforming according to the vector representation of spin and ($a,b,\ldots$) will denote indices transforming according to the adjoint representation of the $SU(N_f)$ flavor group.

Spin projection operators are easily adapted from Eqs.~(\ref{eq:p0s}), (\ref{eq:p1s}), and (\ref{eq:p3s}) as
\begin{equation}
\left[ \pr{J=0}_\mathrm{spin} \right]^{j_1j_2k_1k_2} = \frac13 \delta^{j_1j_2} \delta^{k_1k_2}, \label{eq:j0}
\end{equation}
\begin{equation}
\left[ \pr{J=1}_\mathrm{spin} \right]^{j_1j_2k_1k_2} = \frac12 (\delta^{j_1k_1} \delta^{j_2k_2} - \delta^{j_2k_1} \delta^{j_1k_2}), \label{eq:j1}
\end{equation}
\begin{equation}
\left[ \pr{J=2}_\mathrm{spin} \right]^{j_1j_2k_1k_2} = \frac12 (\delta^{j_1k_1} \delta^{j_2k_2} + \delta^{j_2k_1} \delta^{j_1k_2}) - \frac13 \delta^{k_1k_2} \delta^{j_1j_2}. \label{eq:j2}
\end{equation}

As for flavor projection operators, the tensor product of two adjoints can be separated into an antisymmetric and a symmetric product, $(adj \otimes adj)_A$ and $(adj \otimes adj)_S$, respectively. In the notation of Ref.~\cite{djm95}, these products are written as
\begin{subequations}
\begin{equation}
(adj \otimes adj)_A = adj \oplus \bar{a}s \oplus \bar{s}a,
\end{equation}
and
\begin{equation}
(adj \otimes adj)_S = 1 \oplus adj \oplus \bar{s}s \oplus \bar{a}a.
\end{equation}
\end{subequations}

Thus, the explicit forms of flavor projection operators read as
\begin{equation}
\left[ \pr{1}_\mathrm{flavor} \right]^{a_1a_2b_1b_2} = \frac{1}{N_f^2-1} \delta^{a_1a_2} \delta^{b_1b_2}, \label{eq:fs}
\end{equation}
\begin{equation}
\left[ \pr{adj}_\mathrm{flavor} \right]^{a_1a_2b_1b_2} = \frac{1}{N_f} f^{a_1a_2c} f^{b_1b_2c} + \frac{N_f}{N_f^2-4} d^{a_1a_2c} d^{b_1b_2c}, \label{eq:fo}
\end{equation}
\begin{equation}
\left[ \pr{\bar{a}s+\bar{s}a}_\mathrm{flavor} \right]^{a_1a_2b_1b_2} = \frac12 (\delta^{a_1b_1} \delta^{a_2b_2} - \delta^{a_2b_1} \delta^{a_1b_2}) - \frac{1}{N_f} f^{a_1a_2c}f^{b_1b_2c}, \label{eq:ft}
\end{equation}
\begin{eqnarray}
\left[ \pr{\bar{s}s}_\mathrm{flavor} \right]^{a_1a_2b_1b_2} & = & \frac{N_f+2}{4N_f} (\delta^{a_1b_1} \delta^{a_2b_2} + \delta^{a_2b_1} \delta^{a_1b_2}) - \frac{N_f+2}{2N_f(N_f+1)} \delta^{a_1a_2} \delta^{b_1b_2} - \frac{N_f+4}{4(N_f+2)} d^{a_1a_2c}d^{b_1b_2c} \nonumber \\
&  & \mbox{} + \frac14 (d^{a_1b_1c} d^{a_2b_2c} + d^{a_2b_1c} d^{a_1b_2c}), \label{eq:aa}
\end{eqnarray}
\begin{eqnarray}
\left[ \pr{\bar{a}a}_\mathrm{flavor} \right]^{a_1a_2b_1b_2} & = & \frac{N_f-2}{4N_f} (\delta^{a_1b_1} \delta^{a_2b_2} + \delta^{a_2b_1} \delta^{a_1b_2}) + \frac{N_f-2}{2N_f(N_f-1)} \delta^{a_1a_2} \delta^{b_1b_2} + \frac{N_f-4}{4(N_f-2)} d^{a_1a_2c} d^{b_1b_2c} \nonumber \\
&  & \mbox{} - \frac14 (d^{a_1b_1c} d^{a_2b_2c} + d^{a_2b_1c} d^{a_1b_2c}). \label{eq:ss}
\end{eqnarray}
It should be remarked that the first and second summands of Eq.~(\ref{eq:fo}) define the antisymmetric and symmetric components of $\left[ \pr{adj}_\mathrm{flavor} \right]^{a_1a_2b_1b_2}$, respectively.

Let us also notice that
\begin{equation}
\left[ \pr{\bar{s}s}_\mathrm{flavor} + \pr{\bar{a}a}_\mathrm{flavor} \right]^{a_1a_2b_1b_2} = \frac12 (\delta^{a_1b_1} \delta^{a_2b_2} + 
\delta^{a_1b_2} \delta^{a_2b_1}) - \frac{1}{N_f^2-1} \delta^{a_1a_2} \delta^{b_1b_2} - \frac{N_f}{N_f^2-4} d^{a_1a_2c} d^{b_1b_2c}. \label{eq:s27}
\end{equation}

Implicit forms of the projectors (\ref{eq:ft}) and (\ref{eq:s27}) can be inferred, respectively, from Eqs.~(A13) and (A17) of Ref.~\cite{djm95}. Both approaches yield the same results.

\subsection{Applications of spin and flavor projection operators in the $1/N_c$ operator expansion}

The way spin and flavor projection operators work can be better seen through a few examples. For definiteness, the analysis can be confined 
to the physically interesting case of $N_f=3$ light quark flavors; thus, the lowest-lying baryon states fall into a representation of the $SU(6)$ spin-flavor group, which decomposes as $SU(2)\otimes SU(3)$.

For the $SU(3)$ flavor group, the $adj$, $\bar{a}s+\bar{s}a$, and $\bar{s}s$ representations are the $8$, $10+\overline{10}$, and $27$, respectively, while the representation $\bar{a}a$ does not exist. In consequence, it can be shown that
\begin{equation}
\left[\pr{\bar{a}a}_\mathrm{flavor} Q_1Q_2\right]^{a_1a_2} = 0
\end{equation}
for $SU(3)$.

First, let us analyze the two-body operator $J^{j_1}J^{j_2}$, which is a spin-2 object. It can be written as
\begin{equation}
J^{j_1}J^{j_2} = \frac12 \{J^{j_1},J^{j_2}\} + \frac12 [J^{j_1},J^{j_2}]. \label{eq:spr}
\end{equation}

Projecting out the $J=0$, $J=1$, and $J=2$ components of this product of operators is straightforwardly done with the help of projection operators (\ref{eq:j0}), (\ref{eq:j1}), and (\ref{eq:j2}). The spin projections for the operator $J^{j_1}J^{j_2}$ read 
\begin{subequations}
\begin{equation}
\left[ \pr{J=0}_\mathrm{spin} \right]^{k_1k_2j_1j_2} (J^{j_1}J^{j_2}) = \frac13 \delta^{k_1k_2}J^2,
\end{equation}
\begin{equation}
\left[ \pr{J=1}_\mathrm{spin} \right]^{k_1k_2j_1j_2} (J^{j_1}J^{j_2}) = \frac{i}{2} \epsilon^{k_1k_2i} J^i, 
\end{equation}
and
\begin{equation}
\left[ \pr{J=2}_\mathrm{spin} \right]^{k_1k_2j_1j_2} (J^{j_1}J^{j_2}) = \frac12 \{J^{k_1},J^{k_2}\} - \frac13 \delta^{k_1k_2}J^2.
\end{equation}
\end{subequations}
whereas the nonzero spin projections of the anticommutator and commutator in Eq.~(\ref{eq:spr}) read
\begin{subequations}
\begin{equation}
\left[ \pr{J=0}_\mathrm{spin} \right]^{k_1k_2j_1j_2} \{J^{j_1}, J^{j_2}\} = \frac23 \delta^{k_1k_2}J^2,
\end{equation}
\begin{equation}
\left[ \pr{J=2}_\mathrm{spin} \right]^{k_1k_2j_1j_2} \{J^{j_1}, J^{j_2}\} = \{J^{k_1}, J^{k_2}\} - \frac23 \delta^{k_1k_2}J^2,
\end{equation}
and
\begin{equation}
\left[ \pr{J=1}_\mathrm{spin} \right]^{k_1k_2j_1j_2} [J^{j_1}, J^{j_2}] = i\epsilon^{k_1k_2i}J^i,
\end{equation}
\end{subequations}
where $J^2 \equiv J^iJ^i$. The consistency between these relations can be checked by a simple inspection.

Less trivial examples are found when spin and flavor are simultaneously involved so the corresponding projectors can act in conjunction. For example, the operator $X^{(j_1b_1)(j_2b_2)} = \{G^{j_1b_1},G^{j_2b_2}\} + \{G^{j_2b_1},G^{j_1b_2}\}$ is a spin-2 object and transforms as a flavor $27$. Projecting out the spin $J=0$, $J=1$, and $J=2$ components of this operator yields
\begin{equation}
\left[ \pr{J=0}_\mathrm{spin} \right]^{k_1k_2j_1j_2} (\{G^{j_1b_1},G^{j_2b_2}\} + \{G^{j_1b_2},G^{j_2b_1}\}) = \frac23 \delta^{k_1k_2} \{G^{ib_1},G^{ib_2}\},
\end{equation}
\begin{equation}
\left[ \pr{J=1}_\mathrm{spin} \right]^{k_1k_2j_1j_2} (\{G^{j_1b_1},G^{j_2b_2}\} + \{G^{j_1b_2},G^{j_2b_1}\}) = 0,
\end{equation}
and
\begin{equation}
\left[ \pr{J=2}_\mathrm{spin} \right]^{k_1k_2j_1j_2} (\{G^{j_1b_1},G^{j_2b_2}\} + \{G^{j_1b_2},G^{j_2b_1}\}) = \{G^{k_1b_1},G^{k_2b_2}\} + \{G^{k_1b_2},G^{k_2b_1}\} - \frac23 \delta^{k_1k_2} \{G^{ib_1},G^{ib_2}\}.
\end{equation}

Now, the flavor $1$, $8$, $10+\overline{10}$, and $27$ components of $X^{(j_1b_1)(j_2b_2)}$, for each spin, can be straightforwardly projected out. The $J=0$ projections read
\begin{equation}
\left[ \pr{J=0}_\mathrm{spin} \right]^{k_1k_2j_1j_2} \left[\pr{1}_\mathrm{flavor} \right]^{a_1a_2b_1b_2} (\{G^{j_1b_1},G^{j_2b_2}\} + \{G^{j_1b_2},G^{j_2b_1}\}) = \frac{1}{12} \delta^{k_1k_2} \delta^{a_1a_2} \{G^{ic},G^{ic}\},
\end{equation}
\begin{equation}
\left[ \pr{J=0}_\mathrm{spin} \right]^{k_1k_2j_1j_2} \left[\pr{8}_\mathrm{flavor} \right]^{a_1a_2b_1b_2} (\{G^{j_1b_1},G^{j_2b_2}\} + \{G^{j_1b_2},G^{j_2b_1}\}) = \frac25 \delta^{k_1k_2} d^{a_1a_2c} d^{b_1b_2c} \{G^{ib_1},G^{ib_2}\},
\end{equation}
\begin{equation}
\left[ \pr{J=0}_\mathrm{spin} \right]^{k_1k_2j_1j_2} \left[ \pr{10+\overline{10}}_\mathrm{flavor} \right]^{a_1a_2b_1b_2}(\{G^{j_1b_1},G^{j_2b_2}\} + \{G^{j_1b_2},G^{j_2b_1}\}) = 0,
\end{equation}
\begin{eqnarray}
&  & \left[ \pr{J=0}_\mathrm{spin} \right]^{k_1k_2j_1j_2} \left[ \pr{27}_\mathrm{flavor} \right]^{a_1a_2b_1b_2} (\{G^{j_1b_1},G^{j_2b_2}\} + \{G^{j_1b_2},G^{j_2b_1}\}) = \frac23 \Big[ \delta^{k_1k_2} \{G^{ia_1},G^{ia_2}\} \nonumber \\
&  & \mbox{} - \frac18 \delta^{k_1k_2} \delta^{a_1a_2} \{G^{ic},G^{ic}\} - \frac35 \delta^{k_1k_2} d^{a_1a_2c} d^{b_1b_2c} \{G^{ib_1},G^{ib_2}\} \Big],
\end{eqnarray}
the $J=1$ projections vanish, and the $J=2$ projections become
\begin{eqnarray}
&  & \left[ \pr{J=2}_\mathrm{spin} \right]^{k_1k_2j_1j_2} \left[\pr{1}_\mathrm{flavor} \right]^{a_1a_2b_1b_2} (\{G^{j_1b_1},G^{j_2b_2}\} + \{G^{j_1b_2},G^{j_2b_1}\}) = \nonumber \\
&  & \mbox{} \frac18 \delta^{a_1a_2} \left[ \{G^{k_1b_2},G^{k_2b_2}\} + \{G^{k_2c},G^{k_1c}\} - \frac23 \delta^{k_1k_2} \{G^{ic},G^{ic}\} \right],
\end{eqnarray}
\begin{eqnarray}
&  & \left[ \pr{J=2}_\mathrm{spin} \right]^{k_1k_2j_1j_2} \left[\pr{8}_\mathrm{flavor} \right]^{a_1a_2b_1b_2} (\{G^{j_1b_1},G^{j_2b_2}\} + \{G^{j_1b_2},G^{j_2b_1}\}) = \nonumber \\
&  & \mbox{} \frac35 d^{a_1a_2c} d^{b_1b_2c} \left[ \{G^{k_1b_1},G^{k_2b_2}\} + \{G^{k_1b_2},G^{k_2b_1}\} - \frac23 \delta^{k_1k_2} \{G^{ib_1},G^{ib_2}\} \right],
\end{eqnarray}
\begin{equation}
\left[ \pr{J=2}_\mathrm{spin} \right]^{k_1k_2j_1j_2} \left[ \pr{10+\overline{10}}_\mathrm{flavor} \right]^{a_1a_2b_1b_2} (\{G^{j_1b_1},G^{j_2b_2}\}+\{G^{j_1b_2},G^{j_2b_1}\}) = 0,
\end{equation}
\begin{eqnarray}
&  & \left[ \pr{J=2}_\mathrm{spin} \right]^{k_1k_2j_1j_2} \left[ \pr{27}_\mathrm{flavor} \right]^{a_1a_2b_1b_2} (\{G^{j_1b_1},G^{j_2b_2}\} + \{G^{j_1b_2},G^{j_2b_1}\}) = \{G^{k_1a_1},G^{k_2a_2}\} + \{G^{k_1a_2},G^{k_2a_1}\} \nonumber \\
&  & \mbox{} - \frac23 \delta^{k_1k_2} \{G^{ia_1},G^{ia_2}\} - \frac14 \delta^{a_1a_2} \left[ \{G^{k_1c},G^{k_2c}\} - \frac13 \delta^{k_1k_2} \{G^{ic},G^{ic}\} \right] \nonumber \\
&  & \mbox{} - \frac35 d^{a_1a_2c} d^{b_1b_2c} \left[ \{G^{k_1b_1},G^{k_2b_2}\} + \{G^{k_1b_2},G^{k_2b_1}\} - \frac23
\delta^{k_1k_2} \{G^{ib_1},G^{ib_2}\} \right], \label{eq:p2sf}
\end{eqnarray}
In particular, both sides of Eq.~(\ref{eq:p2sf}) are spin-2 objects and transform purely as flavor $27$ tensors, i.e., their spin 0 and 1 components and their flavor singlet and octet components have been properly subtracted off by using the appropriate spin and flavor projectors. Operators of this kind appear in the analysis of baryon quadrupole moments \cite{banda20}.

\section{\label{sec:3adj}Projection operators extended to the tensor space $adj \otimes adj \otimes adj$: A few examples}

Projection operators defined in the tensor space $adj \otimes adj \otimes adj$ can be obtained by extending the approach used in the construction of the corresponding ones in the tensor space $adj \otimes adj$. The starting point is the decomposition of the tensor product $adj \otimes adj$ into the irreps indicated in Eq.~(\ref{eq:adjd}), so the tensor product of the adjoint representation and each of these irreps can be evaluated.

The simplest construction is the tensor product of the adjoint and the singlet representation 1, i.e., $1 \otimes adj = adj$. Therefore, the projector
\begin{equation}
\left[ \pr{adj} \right]^{a_1a_2a_3b_1b_2b_3} = \frac{1}{N^2-1} \delta^{a_1a_2} \delta^{b_1b_2} \delta^{b_3a_3},
\end{equation}
acting on the tensor operator $Q_{1}^{b_1}Q_{2}^{b_2}Q_{3}^{b_3}$ yields
\begin{equation}
\frac{1}{N^2-1} \delta^{a_1a_2} Q_{1}^{e}Q_{2}^{e}Q_{3}^{a_3},
\end{equation}
which transforms as an adjoint operator.

Increasing complexity can be found in the tensor product $\bar{s}s \otimes adj$, which can be represented by
\begin{eqnarray*}
&  & \ytableausetup{mathmode, boxsize=1.8em}
\begin{ytableau}
\scriptstyle 1 & \scriptstyle 2 & \scriptstyle 3 & \scriptstyle 4 \\
\scriptstyle 2 & \scriptstyle 3 \\
\scriptstyle 3 & \scriptstyle 4 \\
\none[\vdots] \\
\scriptstyle N - 1 & \scriptstyle N
\end{ytableau}
\,\, \otimes \,\, adj \,\, = \,\,
\begin{ytableau}
\scriptstyle 1 & \scriptstyle 2 & \scriptstyle 3 & \scriptstyle 4 & \scriptstyle 5 & \scriptstyle 6 \\
\scriptstyle 2 & \scriptstyle 3 & \scriptstyle 4 \\
\none[\vdots] \\
\scriptstyle N-1 & \scriptstyle N & \scriptstyle N+1
\end{ytableau}
\,\, \oplus \,\,
\begin{ytableau}
\scriptstyle 1 & \scriptstyle 2 & \scriptstyle 3 & \scriptstyle 4 & \scriptstyle 5 \\
\scriptstyle 2 & \scriptstyle 3 \\
\none[\vdots] \\
\scriptstyle N-2 & \scriptstyle N-1 \\
\scriptstyle N-1
\end{ytableau}
\,\, \oplus \,\,
\begin{ytableau}
\scriptstyle 1 & \scriptstyle 2 & \scriptstyle 3 & \scriptstyle 4 & \scriptstyle 5 \\
\scriptstyle 2 & \scriptstyle 3 & \scriptstyle 4 & \scriptstyle 5 \\
\scriptstyle 3 & \scriptstyle 4 & \scriptstyle 5 \\
\none[\vdots] \\
\scriptstyle N-1 & \scriptstyle N & \scriptstyle N+1
\end{ytableau} \\[5mm]
&  & \mbox{\hglue1.0truecm} \oplus \,\, 2 \,\,
\begin{ytableau}
\scriptstyle 1 & \scriptstyle 2 & \scriptstyle 3 & \scriptstyle 4 \\
\scriptstyle 2 & \scriptstyle 3 \\
\none[\vdots] \\
\scriptstyle N-1 & \scriptstyle N
\end{ytableau}
\,\, \oplus \,\,
\begin{ytableau}
\scriptstyle 1 & \scriptstyle 2 & \scriptstyle 3 & \scriptstyle 4 \\
\scriptstyle 2 & \scriptstyle 3 & \scriptstyle 4 \\
\scriptstyle 3 & \scriptstyle 4 & \\
\none[\vdots] \\
\scriptstyle N-2 & \scriptstyle N-1 \\
\scriptstyle N-1
\end{ytableau}
\,\, \oplus \,\,
\begin{ytableau}
\scriptstyle 1 & \scriptstyle 2 & \scriptstyle 3 \\
\scriptstyle 2 \\
\none[\vdots] \\
\scriptstyle N-2 
\end{ytableau}
\,\, \oplus \,\,
\begin{ytableau}
\scriptstyle 1 & \scriptstyle 2 & \scriptstyle 3 \\
\scriptstyle 2 & \scriptstyle 3 & \scriptstyle 4 \\
\scriptstyle 3 & \scriptstyle 4 \\
\none[\vdots] \\
\scriptstyle N-1 & \scriptstyle N
\end{ytableau}
\,\, \oplus \,\,
\begin{ytableau}
\scriptstyle 1 & \scriptstyle 2 \\
\scriptstyle 2 \\
\none[\vdots] \\
\scriptstyle N-1
\end{ytableau}
\end{eqnarray*}

\begin{table}[h!]
\centering
\begin{tabular}{|c|c|c|c|}
\hline
Eigenvalue & Representation & Eigenvalue & Representation \\
\hline &  & & \\
$\,\,c_0 = 3(N+2)\,\,$ &
\footnotesize
\ytableausetup{mathmode, boxsize=2.1em}
\begin{ytableau}
1 & 2 & 3 & 4 & 5 & 6 \\
2 & 3 & 4 \\
\none[\vdots] \\
\scriptstyle N-1 & \scriptstyle N & \scriptstyle N+1
\end{ytableau}
& $\,\,c_3 = 3N+2\,\,$ &
\footnotesize
\ytableausetup{mathmode, boxsize=2.1em}
\begin{ytableau}
1 & 2 & 3 & 4 \\
2 & 3 & 4 \\
3 & 4 \\
\none[\vdots] \\
\scriptstyle N - 2 & \scriptstyle N-1 \\
\scriptstyle N-1
\end{ytableau}
\\
& & & \\
\hline &  & & \\
$c_1 = 3(N+1)$ & \,\,
\footnotesize
\ytableausetup{mathmode, boxsize=2.1em}
\begin{ytableau}
1 & 2 & 3 & 4 & 5 \\
2 & 3 \\
\none[\vdots] \\
\scriptstyle N - 2 & \scriptstyle N-1 \\
\scriptstyle N-1
\end{ytableau}
\,$\oplus$
\footnotesize
\ytableausetup{mathmode, boxsize=2.1em}
\begin{ytableau}
1 & 2 & 3 & 4 & 5 \\
2 & 3 & 4 & 5 \\
3 & 4 & 5 \\
\none[\vdots] \\
\scriptstyle N-1 & \scriptstyle N & \scriptstyle N+1
\end{ytableau}
\,\,
& $c_4 = 2 N$ &
\,\,
\footnotesize
\ytableausetup{mathmode, boxsize=2.1em}
\begin{ytableau}
1 & 2 & 3 \\
2 \\
\none[\vdots] \\
\scriptstyle N-2
\end{ytableau}
\,$\oplus$
\footnotesize
\ytableausetup{mathmode, boxsize=2.1em}
\begin{ytableau}
1 & 2 & 3 \\
2 & 3 & 4 \\
3 & 4 \\
\none[\vdots] \\
\scriptstyle N-1 & \scriptstyle N
\end{ytableau}
\,\, \\
& & & \\ \hline & & & \\
$c_2 = 2(N+1)$ &
\footnotesize
\ytableausetup{mathmode, boxsize=2.1em}
\begin{ytableau}
1 & 2 & 3 & 4 \\
2 & 3 \\
\none[\vdots] \\
\scriptstyle N-1 & \scriptstyle N
\end{ytableau}
& $c_5 = N$ &
\footnotesize
\ytableausetup{mathmode, boxsize=2.1em}
\begin{ytableau}
1 & 2 \\
2 \\
\none[\vdots] \\
\scriptstyle N-1
\end{ytableau}
\\
& & & \\
\hline
\end{tabular}
\caption{\label{tab:t3adj}Quadratic Casimir eigenvalues for the representations obtained in the tensor product $\bar{s}s\otimes adj$.}
\end{table}

Let $T^a_{3A}$ denote the generators for the tensor product space $\bar{s}s\otimes adj$. These generators are given by
\begin{equation}
T^a_{3A} = \pr{3} T_{2A}^a \otimes \openone + \pr{3} \otimes T_A^a, \label{eq:generators_3}
\end{equation}
where $T_{2A}^a$ are defined in Eq.~(\ref{eq:T2A_t}). Accordingly, the quadratic Casimir operator reads
\begin{equation}
C = \pr{3} T_{2A}^a T_{2A}^a \otimes \openone + 2 \pr{3} T_{2A}^a \otimes T_A^a + \pr{3} \otimes T_A^a T_A^a,
\end{equation}
whose explicit form in components becomes
\begin{equation}
\left[ C \right]^{a_1a_2a_3b_1b_2b_3} = (3N+2) \left[ \pr{3} \right]^{a_1a_2b_1b_2} \delta^{a_3b_3} - 2 \left\{ \left[ \pr{3} \right]^{a_1a_2b_2e} F^{eb_1a_3b_3} + (b_1 \leftrightarrow b_2) \right\}, \label{eq:Casimir_3}
\end{equation}
which follows from the use of the identity
\begin{equation}
\left[ \pr{3} \right]^{a_1a_2d_1d_2} F^{d_1b_1d_2b_2} = - \left[ \pr{3} \right]^{a_1a_2b_1b_2},
\end{equation}
along with Eqs.~(\ref{eq:casimir_1}) and (\ref{eq:C_c}).

The eigenvalues of the quadratic Casimir operator for each representation are displayed in Table \ref{tab:t3adj}. Following relation (\ref{eq:general_proj}) and gathering together partial results, the corresponding projection operators are
\begin{eqnarray}
\prt{m} & = & \prod_{i=1}^5 \left[ \frac{C-c_{n_i}}{c_m-c_{n_i}} \right] \nonumber \\
& = & \mbox{} \frac{-\tilde{\alpha}_0 + \tilde{\alpha}_1 C - \tilde{\alpha}_2 C^2 + \tilde{\alpha}_3 C^3 - \tilde{\alpha}_4 C^4 + C^5}{\prod_{i=1}^5(c_m-c_{n_i})}, \label{eq:prm888}
\end{eqnarray}
where the coefficients $\tilde{\alpha}_i$ read
\begin{subequations}
\begin{equation}
\tilde{\alpha}_0 = c_{n_1} c_{n_2} c_{n_3} c_{n_4} c_{n_5},
\end{equation}
\begin{equation}
\tilde{\alpha}_1 = c_{n_1} c_{n_2} c_{n_3} c_{n_4} + c_{n_1} c_{n_2} c_{n_3} c_{n_5} + c_{n_1} c_{n_2} c_{n_4} c_{n_5} + c_{n_1} c_{n_3} c_{n_4} c_{n_5} + c_{n_2} c_{n_3} c_{n_4} c_{n_5},
\end{equation}
\begin{eqnarray}
\tilde{\alpha}_2 & = & c_{n_1} c_{n_2} c_{n_3} + c_{n_1} c_{n_2} c_{n_4} + c_{n_1} c_{n_3} c_{n_4} + c_{n_2} c_{n_3} c_{n_4} + c_{n_1} c_{n_2} c_{n_5} + c_{n_1} c_{n_3} c_{n_5} + c_{n_2} c_{n_3} c_{n_5} \nonumber \\
&  & \mbox{} + c_{n_1} c_{n_4} c_{n_5} + c_{n_2} c_{n_4} c_{n_5} + c_{n_3} c_{n_4} c_{n_5},
\end{eqnarray}
\begin{equation}
\tilde{\alpha}_3 = c_{n_1} c_{n_2} + c_{n_1} c_{n_3} + c_{n_2} c_{n_3} + c_{n_1} c_{n_4} + c_{n_2} c_{n_4} + c_{n_3} c_{n_4} + c_{n_1} c_{n_5} + c_{n_2} c_{n_5} + c_{n_3} c_{n_5} + c_{n_4} c_{n_5},
\end{equation}
and
\begin{equation}
\tilde{\alpha}_4 = c_{n_1} + c_{n_2} + c_{n_3} + c_{n_4} + c_{n_5}.
\end{equation}
\end{subequations}

The powers of $C$ required in Eq.~(\ref{eq:prm888}) are obtained as
\begin{subequations}
\begin{equation}
C = a E_0 + 2 E_1, \label{eq:C31}
\end{equation}
\begin{equation}
C^2 = a^2 E_0 + 4 a E_1 + 4 E_2, \label{eq:C32}
\end{equation}
\begin{equation}
C^3 = a^3 E_0 + 6 a^2 E_1 + 12 a E_2 + 8 E_3, \label{eq:C33}
\end{equation}
\begin{equation}
C^4 = a^4 E_0 + 8 a^3 E_1 + 24 a^2 E_2 + 32 a E_3 + 16 E_4, \label{eq:C34}
\end{equation}
and
\begin{equation}
C^5 = a^5 E_0 + 10 a^4 E_1 + 40 a^3 E_2 + 80 a^2 E_3 + 80 a E_4 + 32 E_5, \label{eq:C35}
\end{equation}
\end{subequations}
with $a=3N+2$ and
\begin{subequations}
\begin{equation}
\left[ E_0 \right]^{a_1a_2a_3b_1b_2b_3} = \left[ \pr{3} \right]^{a_1a_2b_1b_2} \delta^{a_3b_3}, \label{eq:E0}
\end{equation}
\begin{equation}
\left[ E_1 \right]^{a_1a_2a_3b_1b_2b_3} = \left[ \pr{3} \right]^{a_1a_2d_1d_2} \left[ T \right]^{d_1d_2a_3b_1b_2b_3}, \label{eq:E1}
\end{equation}
\begin{equation}
\left[ E_2 \right]^{a_1a_2a_3b_1b_2b_3} = \left[ \pr{3} \right]^{a_1a_2d_1d_2} \left[ T^2 \right]^{d_1d_2a_3b_1b_2b_3}, \label{eq:E2}
\end{equation}
\begin{equation}
\left[ E_3 \right]^{a_1a_2a_3b_1b_2b_3} = \left[ \pr{3} \right]^{a_1a_2d_1d_2} \left[ T^3 \right]^{d_1d_2a_3b_1b_2b_3}, \label{eq:E3}
\end{equation}
\begin{equation}
\left[ E_4 \right]^{a_1a_2a_3b_1b_2b_3} = \left[ \pr{3} \right]^{a_1a_2d_1d_2} \left[ T^4 \right]^{d_1d_2a_3b_1b_2b_3}, \label{eq:E4}
\end{equation}
and
\begin{equation}
\left[ E_5 \right]^{a_1a_2a_3b_1b_2b_3} = \left[ \pr{3} \right]^{a_1a_2d_1d_2} \left[ T^5 \right]^{d_1d_2a_3b_1b_2b_3}. \label{eq:E5}
\end{equation}
\end{subequations}
Here $E_0$ represents the identity for the tensor product space under consideration and the tensor $\left[ T \right]^{a_1a_2a_3b_1b_2b_3}$ is defined as
\begin{eqnarray}
\left[ T \right]^{a_1a_2a_3b_1b_2b_3} & = & \left[ T_{2A}^a \otimes T_A^a \right]^{a_1a_2a_3b_1b_2b_3} \nonumber \\
& = & \mbox{} - F^{a_1b_1a_3b_3} \delta^{a_2b_2} - F^{a_2b_2a_3b_3} \delta^{a_1b_1}. \label{eq:t1}
\end{eqnarray}

The final expression for the projectors $\prt{m}$ can be cast into the compact form
\begin{equation}
\prt{m} = \frac{1}{h_m} \left[ e_0^{(m)} E_0 + e_1^{(m)} E_1 + e_2^{(m)} E_2 + e_3^{(m)} E_3 + e_4^{(m)} E_4 + e_5^{(m)} E_5 \right], \label{eq:prt}
\end{equation}
where the coefficients $h_m$ and $e_n^{(m)}$ are listed in Appendix \ref{sec:coef}.

A long and tedious but otherwise standard calculation is required to prove that
\begin{equation}
\left[ \prt{m}\right]^{a_1a_2a_3d_1d_2d_3} \left[ \prt{n} \right]^{d_1d_2d_3b_1b_2b_3} = \left\{
\begin{array}{ll}
0, & \quad m \neq n \\[3mm]
\displaystyle \left[ \prt{m} \right]^{a_1a_2a_3b_1b_2b_3}, & \quad m = n
\end{array}
\right.
\end{equation}
and
\begin{equation}
\sum_{m=0}^5 \left[ \prt{m} \right]^{a_1a_2a_3b_1b_2b_3} = E_0^{a_1a_2a_3b_1b_2b_3}.
\end{equation}

\subsection{An example of projection operators in $SU(2)$}

In this case the projector $\pr{3}$ corresponds to the representation with spin-2 given in Eq.~(\ref{eq:p3s}). Therefore, the projector $\prt{5}$ which, according to Table \ref{tab:t3adj}, corresponds to an adjoint representation (spin-1 with three indices) and is given by
\begin{eqnarray}
\prt{5} & = & \frac{e_0^{(m)} E_0 + e_1^{(5)} E_1 + e_2^{(5)} E_2 + e_3^{(5)} E_3 + e_4^{(5)} E_4 + e_5^{(5)} E_5}{h_5} \nonumber \\
& = & \mbox{} \frac{- 4 E_1 + 4 E_2 + 9 E_3 - E_4 - 2 E_5}{210}.
\end{eqnarray}

Using the expressions for $E_i$ given in (\ref{eq:E0}--\ref{eq:E5}) for $N=2$, $\prt{5}$ in components can be rewritten as
\begin{eqnarray}
\left[ \prt{5} \right]^{a_1a_2a_3b_1b_2b_3} & = & \frac{1}{15} \delta^{a_1a_2} \delta^{b_1b_2} \delta^{a_3b_3} + \frac{3}{20} \left\{ \delta^{a_1a_3}\delta^{a_2b_2} \delta^{b_1b_3} + \delta^{a_1a_3}\delta^{a_2b_1} \delta^{b_2b_3} + (a_1,b_1) \leftrightarrow (a_2,b_2) \right\} \nonumber \\
&  & \mbox{} - \frac{1}{10} \left\{ \delta^{a_3b_2} \delta^{a_1a_2} \delta^{b_1b_3} + \delta^{a_1b_3} \delta^{a_2c_3} \delta^{b_1b_2} + (a_1,b_1) \leftrightarrow (a_2,b_2) \right\}.
\end{eqnarray}

\subsection{An example of projection operators in $SU(3)$}

Formally, given three $SU(3)$ adjoints $Q_1^{a_1}$, $Q_2^{a_2}$, and $Q_3^{a_3}$, the tensor product between them, $Q_1^{a_1}Q_2^{a_2}Q_3^{a_3}$, possesses all flavor $1$, $8$, $10+\overline{10}$, $27$, $35+\overline{35}$, and $64$ components. Operators transforming in the flavor $64$ representation, for instance, are relevant in the analysis of baryon mass splittings of the spin-$1/2$ octet and spin-$3/2$ decuplet baryons in the $1/N_c$ expansion combined with perturbative flavor breaking at order $\mathcal{O}(\epsilon^2)$, where $\epsilon\sim m_s$ is a (dimensionless) measure of $SU(3)$ breaking \cite{Jenkins_mass}.

In order to subtract off all but the flavor $64$ component, the projection operator $\prt{m}$, for $m=0$, is constructed following the lines of Eq.~(\ref{eq:prt}); this procedure leads to $\prt{64}_\mathrm{flavor}$. The eigenvalue of the Casimir operator is $c_0=3(N_f+2)$ and the corresponding Young tableau can easily be obtained from the corresponding one depicted in Table \ref{tab:t3adj} for $N_f=3$.

Let $Q^{(64)}$ be the operator that transforms as a genuine flavor $64$. It is thus given by
\begin{equation}
\left[ Q^{(64)} \right]^{a_1a_2a_3} = \left[ \prt{64}_\mathrm{flavor} Q_1Q_2Q_3 \right]^{a_1a_2a_3}.
\end{equation}
The projection operator $\prt{64}_\mathrm{flavor}$ itself has a rather involved form, containing several hundreds of terms. Because of the length and unilluminating nature of the resultant expression, it is more convenient to list a few components of $Q^{(64)}$. For instance,
\begin{eqnarray}
\left[ Q^{(64)} \right]^{888} & = & \frac{1}{70} \Big[ 3 Q_1^8Q_2^1Q_3^1 - \sqrt{3} Q_1^6Q_2^4Q_3^1 - \sqrt{3} Q_1^7Q_2^5Q_3^1 - \sqrt{3}Q_1^4Q_2^6Q_3^1 - \sqrt{3} Q_1^5Q_2^7Q_3^1 + 3 Q_1^1Q_2^8Q_3^1 \nonumber \\
&  & \mbox{\hglue0.5truecm} + 3 Q_1^8Q_2^2Q_3^2 + \sqrt{3} Q_1^7Q_2^4Q_3^2 - \sqrt{3} Q_1^6Q_2^5Q_3^2 - \sqrt{3} Q_1^5Q_2^6Q_3^2 + \sqrt{3} Q_1^4Q_2^7Q_3^2 + 3 Q_1^2Q_2^8Q_3^2 \nonumber \\
&  & \mbox{\hglue0.5truecm} + 3 Q_1^8Q_2^3Q_3^3 - \sqrt{3} Q_1^4Q_2^4Q_3^3 - \sqrt{3} Q_1^5Q_2^5Q_3^3 + \sqrt{3} Q_1^6Q_2^6Q_3^3 + \sqrt{3} Q_1^7Q_2^7Q_3^3 + 3 Q_1^3Q_2^8Q_3^3 \nonumber \\
&  & \mbox{\hglue0.5truecm} - \sqrt{3} Q_1^6Q_2^1Q_3^4 + \sqrt{3} Q_1^7Q_2^2Q_3^4 - \sqrt{3} Q_1^4Q_2^3Q_3^4 - \sqrt{3} Q_1^3Q_2^4Q_3^4 - 9 Q_1^8Q_2^4Q_3^4 - \sqrt{3} Q_1^1Q_2^6Q_3^4 \nonumber \\
&  & \mbox{\hglue0.5truecm} + \sqrt{3} Q_1^2Q_2^7Q_3^4 - 9 Q_1^4Q_2^8Q_3^4 - \sqrt{3} Q_1^7Q_2^1Q_3^5 - \sqrt{3} Q_1^6Q_2^2Q_3^5 - \sqrt{3} Q_1^5Q_2^3Q_3^5 - \sqrt{3} Q_1^3Q_2^5Q_3^5 \nonumber \\
&  & \mbox{\hglue0.5truecm} - 9 Q_1^8Q_2^5Q_3^5 - \sqrt{3} Q_1^2Q_2^6Q_3^5 - \sqrt{3} Q_1^1Q_2^7Q_3^5 - 9 Q_1^5Q_2^8Q_3^5 - \sqrt{3} Q_1^4Q_2^1Q_3^6 - \sqrt{3} Q_1^5Q_2^2Q_3^6 \nonumber \\
&  & \mbox{\hglue0.5truecm} + \sqrt{3} Q_1^6Q_2^3Q_3^6 - \sqrt{3} Q_1^1Q_2^4Q_3^6 - \sqrt{3} Q_1^2Q_2^5Q_3^6 + \sqrt{3} Q_1^3Q_2^6Q_3^6 - 9 Q_1^8Q_2^6Q_3^6 - 9 Q_1^6Q_2^8Q_3^6 \nonumber \\
&  & \mbox{\hglue0.5truecm} - \sqrt{3} Q_1^5Q_2^1Q_3^7 + \sqrt{3} Q_1^4Q_2^2Q_3^7 + \sqrt{3} Q_1^7Q_2^3Q_3^7 + \sqrt{3} Q_1^2Q_2^4Q_3^7 - \sqrt{3} Q_1^1Q_2^5Q_3^7 + \sqrt{3} Q_1^3Q_2^7Q_3^7 \nonumber \\
&  & \mbox{\hglue0.5truecm} - 9 Q_1^8Q_2^7Q_3^7 - 9 Q_1^7Q_2^8Q_3^7 + 3 Q_1^1Q_2^1Q_3^8 + 3 Q_1^2Q_2^2Q_3^8 + 3 Q_1^3Q_2^3Q_3^8 - 9 Q_1^4Q_2^4Q_3^8 \nonumber \\
&  & \mbox{\hglue0.5truecm} - 9 Q_1^5Q_2^5Q_3^8 - 9 Q_1^6Q_2^6Q_3^8 - 9 Q_1^7Q_2^7Q_3^8 + 27 Q_1^8Q_2^8Q_3^8 \Big], \label{eq:q888}
\end{eqnarray}
\begin{eqnarray}
\left[Q^{(64)} \right]^{883} & = & \frac{1}{126} \Big[ Q_1^3Q_2^1Q_3^1 + Q_1^1Q_2^3Q_3^1 + Q_1^3Q_2^2Q_3^2 + Q_1^2Q_2^3Q_3^2 + Q_1^1Q_2^1Q_3^3 + Q_1^2Q_2^2Q_3^3 + 3 Q_1^3Q_2^3Q_3^3 \nonumber \\
&  & \mbox{\hglue0.8truecm} - 5 Q_1^4Q_2^4Q_3^3 - 5 Q_1^5Q_2^5Q_3^3 - 5 Q_1^6Q_2^6Q_3^3 - 5 Q_1^7Q_2^7Q_3^3 + 15 Q_1^8Q_2^8Q_3^3 - 5 Q_1^4Q_2^3Q_3^4 \nonumber \\
&  & \mbox{\hglue0.8truecm} - 5 Q_1^3Q_2^4Q_3^4 - 5\sqrt{3} Q_1^8Q_2^4Q_3^4 - 5\sqrt{3} Q_1^4Q_2^8Q_3^4 - 5 Q_1^5Q_2^3Q_3^5 - 5 Q_1^3Q_2^5Q_3^5 - 5\sqrt{3} Q_1^8Q_2^5Q_3^5 \nonumber \\
&  & \mbox{\hglue0.8truecm} - 5\sqrt{3} Q_1^5Q_2^8Q_3^5 - 5 Q_1^6Q_2^3Q_3^6 - 5 Q_1^3Q_2^6Q_3^6 + 5\sqrt{3} Q_1^8Q_2^6Q_3^6 + 5\sqrt{3} Q_1^6Q_2^8Q_3^6 - 5 Q_1^7Q_2^3Q_3^7 \nonumber \\
&  & \mbox{\hglue0.8truecm} - 5 Q_1^3Q_2^7Q_3^7 + 5\sqrt{3} Q_1^8Q_2^7Q_3^7 + 5\sqrt{3} Q_1^7Q_2^8Q_3^7 + 15 Q_1^8Q_2^3Q_3^8 - 5\sqrt{3} Q_1^4Q_2^4Q_3^8 \nonumber \\
&  & \mbox{\hglue0.8truecm} - 5\sqrt{3} Q_1^5Q_2^5Q_3^8 + 5\sqrt{3} Q_1^6Q_2^6Q_3^8 + 5\sqrt{3} Q_1^7Q_2^7Q_3^8 + 15 Q_1^3Q_2^8Q_3^8 \Big],
\end{eqnarray}
\begin{eqnarray}
\left[ Q^{(64)} \right]^{833} & = & \frac{1}{70\sqrt{3}} \Big[ - 3\sqrt{3} Q_1^8Q_2^1Q_3^1 + 3 Q_1^6Q_2^4Q_3^1 + 3 Q_1^7Q_2^5Q_3^1 + 3 Q_1^4Q_2^6Q_3^1 + 3 Q_1^5Q_2^7Q_3^1 - 3\sqrt{3} Q_1^1Q_2^8Q_3^1 \nonumber \\
&  & \mbox{\hglue1.15truecm} - 3\sqrt{3} Q_1^8Q_2^2Q_3^2 - 3 Q_1^7Q_2^4Q_3^2 + 3 Q_1^6Q_2^5Q_3^2 + 3 Q_1^5Q_2^6Q_3^2 - 3 Q_1^4Q_2^7Q_3^2 - 3\sqrt{3} Q_1^2Q_2^8Q_3^2 \nonumber \\
&  & \mbox{\hglue1.15truecm} + 7\sqrt{3} Q_1^8Q_2^3Q_3^3 - 7 Q_1^4Q_2^4Q_3^3 - 7 Q_1^5Q_2^5Q_3^3 + 7 Q_1^6Q_2^6Q_3^3 + 7 Q_1^7Q_2^7Q_3^3 + 7\sqrt{3} Q_1^3Q_2^8Q_3^3 \nonumber \\
&  & \mbox{\hglue1.15truecm} + 3 Q_1^6Q_2^1Q_3^4 - 3 Q_1^7Q_2^2Q_3^4 - 7 Q_1^4Q_2^3Q_3^4 - 7 Q_1^3Q_2^4Q_3^4 - \sqrt{3} Q_1^8Q_2^4Q_3^4 + 3 Q_1^1Q_2^6Q_3^4 \nonumber \\
&  & \mbox{\hglue1.15truecm} - 3 Q_1^2Q_2^7Q_3^4 - \sqrt{3} Q_1^4Q_2^8Q_3^4 + 3 Q_1^7Q_2^1Q_3^5 + 3 Q_1^6Q_2^2Q_3^5 - 7 Q_1^5Q_2^3Q_3^5 - 7 Q_1^3Q_2^5Q_3^5 \nonumber \\
&  & \mbox{\hglue1.15truecm} - \sqrt{3} Q_1^8Q_2^5Q_3^5 + 3 Q_1^2Q_2^6Q_3^5 + 3 Q_1^1Q_2^7Q_3^5 - \sqrt{3} Q_1^5Q_2^8Q_3^5 + 3 Q_1^4Q_2^1Q_3^6 + 3 Q_1^5Q_2^2Q_3^6 \nonumber \\
&  & \mbox{\hglue1.15truecm} + 7 Q_1^6Q_2^3Q_3^6 + 3 Q_1^1Q_2^4Q_3^6 + 3 Q_1^2Q_2^5Q_3^6 + 7 Q_1^3Q_2^6Q_3^6 - \sqrt{3} Q_1^8Q_2^6Q_3^6 - \sqrt{3} Q_1^6Q_2^8Q_3^6 \nonumber \\
&  & \mbox{\hglue1.15truecm} + 3 Q_1^5Q_2^1Q_3^7 - 3 Q_1^4Q_2^2Q_3^7 + 7 Q_1^7Q_2^3Q_3^7 - 3 Q_1^2Q_2^4Q_3^7 + 3 Q_1^1Q_2^5Q_3^7 + 7 Q_1^3Q_2^7Q_3^7 \nonumber \\
&  & \mbox{\hglue1.15truecm} - \sqrt{3} Q_1^8Q_2^7Q_3^7 - \sqrt{3} Q_1^7Q_2^8Q_3^7 - 3\sqrt{3} Q_1^1Q_2^1Q_3^8 - 3\sqrt{3} Q_1^2Q_2^2Q_3^8 + 7\sqrt{3} Q_1^3Q_2^3Q_3^8 \nonumber \\
&  & \mbox{\hglue1.15truecm} - \sqrt{3} Q_1^4Q_2^4Q_3^8 - \sqrt{3} Q_1^5Q_2^5Q_3^8 - \sqrt{3} Q_1^6Q_2^6Q_3^8 - \sqrt{3} Q_1^7Q_2^7Q_3^8 + 3\sqrt{3} Q_1^8Q_2^8Q_3^8 \Big],
\end{eqnarray}
and
\begin{eqnarray}
\left[Q^{(64)} \right]^{333} & = & \frac{1}{126} \Big[ - 25 Q_1^3Q_2^1Q_3^1 - 25 Q_1^1Q_2^3Q_3^1 - 25 Q_1^3Q_2^2Q_3^2 - 25 Q_1^2Q_2^3Q_3^2 - 25 Q_1^1Q_2^1Q_3^3 - 25 Q_1^2Q_2^2Q_3^3 \nonumber \\
&  & \mbox{\hglue0.8truecm} + 51 Q_1^3Q_2^3Q_3^3 - Q_1^4Q_2^4Q_3^3 - Q_1^5Q_2^5Q_3^3 - Q_1^6Q_2^6Q_3^3 - Q_1^7Q_2^7Q_3^3 + 3 Q_1^8Q_2^8Q_3^3 - Q_1^4Q_2^3Q_3^4 \nonumber \\
&  & \mbox{\hglue0.8truecm} - Q_1^3Q_2^4Q_3^4 - \sqrt{3} Q_1^8Q_2^4Q_3^4 - \sqrt{3} Q_1^4Q_2^8Q_3^4 - Q_1^5Q_2^3Q_3^5 - Q_1^3Q_2^5Q_3^5 - \sqrt{3} Q_1^8Q_2^5Q_3^5 \nonumber \\
&  & \mbox{\hglue0.8truecm} - \sqrt{3} Q_1^5Q_2^8Q_3^5 - Q_1^6Q_2^3Q_3^6 - Q_1^3Q_2^6Q_3^6 + \sqrt{3} Q_1^8Q_2^6Q_3^6 + \sqrt{3} Q_1^6Q_2^8Q_3^6 - Q_1^7Q_2^3Q_3^7 \nonumber \\
&  & \mbox{\hglue0.8truecm} - Q_1^3Q_2^7Q_3^7 + \sqrt{3} Q_1^8Q_2^7Q_3^7 + \sqrt{3} Q_1^7Q_2^8Q_3^7 + 3 Q_1^8Q_2^3Q_3^8 - \sqrt{3} Q_1^4Q_2^4Q_3^8 - \sqrt{3} Q_1^5Q_2^5Q_3^8 \nonumber \\
&  & \mbox{\hglue0.8truecm} + \sqrt{3} Q_1^6Q_2^6Q_3^8 + \sqrt{3} Q_1^7Q_2^7Q_3^8 + 3 Q_1^3Q_2^8Q_3^8 \Big]. \label{eq:q333}
\end{eqnarray}
Note in expressions (\ref{eq:q888}--\ref{eq:q333}) the symmetry under interchange of any two flavor indices, as required for flavor-64 operators.

Returning to the issue of the analysis of baryon mass splittings in the $1/N_c$ expansion combined with perturbative flavor breaking at order $\mathcal{O}(\epsilon^2)$ \cite{Jenkins_mass}, relations (\ref{eq:q888}--\ref{eq:q333}) can be adapted and used in the evaluation of operator structures such as the three-body operator $\{T^a,\{T^b,T^c\}\}$, or even higher-order operators such as $\{T^a,\{T^b,\{J^i,G^{ic}\}\}\}$, $\{T^a,\{\{J^i,G^{ib}\},\{J^j,G^{jc}\}\}\}$, and so on. Therefore, the method introduced here becomes a useful tool to effectively project out spin and flavor representation components in the analysis of large-$N_c$ baryons.

\section{\label{sec:clo}Concluding remarks}

In this paper, the quadratic Casimir operator of the $SU(N)$ group is employed to construct projection operators that can decompose any of its reducible finite-dimensional representation spaces contained in the tensor product of two and three adjoint spaces into irreducible components. The method was first introduced for the Lorentz group in Ref.~\cite{Guzman:2019} and has proven to be quite effective for $SU(N)$.

The projection operators were computed first for the tensor space $adj\otimes adj$. For $N>3$, there are five irreducible representations contained in $adj\otimes adj$, with well-defined eigenvalues of the Casimir operator $C$. This information is summarized in Table \ref{tab:1}. The corresponding projectors are explicitly given in Eqs.~(\ref{eq:p0})-(\ref{eq:p4}). For the tensor space $adj\otimes adj\otimes adj$, the complexity raises considerably, so only the subspace $\bar{s}s\otimes adj$ is studied in detail. This information is summarized in Table \ref{tab:t3adj}. The corresponding projectors are provided in Eq.~(\ref{eq:prt}).

Although the method is general enough, it is specialized to the $SU(2N_f)\to SU(2)\otimes SU(N_f)$ spin-flavor symmetry. The approach thus leads to the construction of spin and flavor projection operators, which can be implemented in the analysis of the $1/N_c$ operator expansion. The use of projection operators allows one to successfully project out the desired components of a given operator and subtract off those that are not needed. To exemplify the method, the projection operators are applied to adjoint tensor operators with two and three flavor indices which, for $SU(3)$, fall into flavor-$27$ and flavor-$64$ representations, respectively. The projectors effectively project out spin and flavor representations of operator structures present in analyses of baryon mass splittings or baryon quadrupole moments, for instance. 

The applicability of the approach is not limited to large-$N_c$ QCD. The approach presented here paves the way to potential applications in shell models of atomic and nuclear physics to construct tensor operators which, with the aid of the Wigner-Eckart theorem, can be used to calculate transition amplitudes. Further applications to the worldline approach to non-Abelian gauge fields should also be seriously considered. In particular, for models that require the construction of a Hamiltonian with an $SU(N)$ symmetry, the method can provide a mechanism to obtain the different irreducible contributions of the operators that appear in such a Hamiltonian. This way the relevance of each different contribution to the spectra can be studied. A clear example can be found in the interacting boson model of nuclear physics \cite{ibm}.

A well-known procedure advocated in the literature to deal with the direct products of irreps of $SU(N)$ (mosty for $N=2$ and $3$) is based on the derivation of Clebsh-Gordan (CG) coefficients, either analytically \cite{deswart,mcnamee} or numerically \cite{alex}. CG coefficients arise in the decomposition of the tensor product of the representation spaces of two irreps of some group into a direct sum of irreducible representation spaces. The utility of CG coefficients in characterizing hadronic decays is irrefutable. States are usually labeled by $|N,Y,I,I_3\rangle$, where $Y$ and $I$ stand for hypercharge and isospin, respectively, and $I_3$ represents the third component of isospin. The method discussed here encodes the information on these coefficients in the components of the projectors, although there is neither an obvious nor a direct relation between them. For example, for $SU(2)$, in order to find the relations that connect the CG coefficients with the projectors, the first step would consist in changing, after projection, the spin generators to a spherical tensor basis. Afterward, the CG coefficients could be found. In applications where there are operators that satisfy Eq.~(\ref{eq:sun_op}), the method presented here has the advantage of working directly with these operators rather than the states mentioned above. In contrast, using CG coefficients requires to change first to a basis where these coefficients are defined. In addition, the projector method gives general expressions in terms of the $f^{abc}$ and $d^{abc}$ symbols, without having to specify a value of $N$.

To close this paper, it should be pointed out that, for a given representation, constructing the whole set of projection operators might seem uninviting for computational difficulty; nonetheless, the technique represents a powerful tool to project out flavor components rigorously.

\begin{acknowledgments}
The authors are grateful to Consejo Nacional de Ciencia y Tecnolog{\'\i}a (Mexico) for support.
\end{acknowledgments}

\appendix

\section{\label{sec:idenSUN} Identities involving structure constants of the Lie algebra of $SU(N)$}

In this section, some relations between the structure constants of the Lie algebra of $SU(N)$ used repeatedly in the present analysis are provided. The list by no means is exhaustive, but it ranges from the Jacobi identity up to the product of 8 $f$'s. The relations read
\begin{equation}
F^{a_1a_2b_1b_2} + F^{b_1a_1a_2b_2} + F^{a_2b_1a_1b_2} = 0, \label{eq:jacobi}
\end{equation}
\begin{equation}
F^{a_1a_2b_1b_2} = \frac{2}{N} (\delta^{a_1b_1} \delta^{a_2b_2} - \delta^{a_1b_2} \delta^{a_2b_1}) + D^{a_1b_1a_2b_2} - D^{a_1b_2a_2b_1}, \label{eq:id_fd}
\end{equation}
\begin{equation}
F^{a_1e_1a_2e_1} = N \delta^{a_1a_2},
\end{equation}
\begin{equation}
F^{a_1a_2e_1e_2} F^{b_1e_1b_2e_2} = \frac{N}{2} F^{a_1a_2b_1b_2},
\end{equation}
\begin{equation}
D^{a_1a_2e_1e_2} F^{b_1e_1b_2e_2} = \frac{N}{2} D^{a_1a_2b_1b_2},
\end{equation}
\begin{eqnarray}
D^{a_1e_1a_2e_2} F^{b_1e_1b_2e_2} & = & \delta^{a_1a_2} \delta^{b_1b_2} - \frac12 \delta^{a_1b_1} \delta^{a_2b_2} - \frac12 \delta^{a_1b_2} \delta^{a_2b_1} + \frac{N}{4} D^{a_1a_2b_1b_2} + \frac{N^2-8}{4N} F^{a_1b_1a_2b_2} \nonumber \\
&  & \mbox{} - \frac{N}{4} F^{a_1b_2a_2b_1},
\end{eqnarray}
\begin{equation}
F^{a_1e_1a_2e_2} F^{b_1e_1b_2e_2} = \delta^{a_1a_2} \delta^{b_1b_2} + \frac12 \delta^{a_1b_1} \delta^{a_2b_2} + \frac12 \delta^{a_1b_2} \delta^{a_2b_1} + \frac{N}{4} (D^{a_1a_2b_1b_2} + F^{a_1a_2b_1b_2}), \label{id_4f}
\end{equation}
\begin{equation}
F^{a_1e_1a_2e_2} F^{e_1e_3e_2e_4} F^{b_1e_3b_2e_4} = N \delta^{a_1a_2} \delta^{b_1b_2} + \frac{N^2}{8} (D^{a_1a_2b_1b_2} + F^{a_1a_2b_1b_2}) + \frac12 (F^{a_1b_1a_2b_2} + F^{a_1b_2a_2b_1}), \label{id_6f}
\end{equation}
\begin{eqnarray}
F^{a_1e_1b_1e_2} F^{a_2e_3b_2e_4} F^{e_3e_5e_4e_6} F^{e_1e_5e_2e_6} & = & \frac{N^2+6}{8} \delta^{a_1a_2} \delta^{b_1b_2} + \frac{7N^2 + 4}{8} \delta^{a_1b_1} \delta^{a_2b_2} + \frac34 \delta^{a_1b_2} \delta^{a_2b_1} \nonumber \\
&  & \mbox{} + \frac{N(N^2+2)}{16} (F^{a_1a_2b_1b_2} + D^{a_1a_2b_1b_2}) + \frac{N}{8} (D^{a_1b_2a_2b_1} - F^{a_1b_2a_2b_1}). \nonumber \\ \label{id_8f}
\end{eqnarray}

\section{\label{sec:coef}Defining coefficients of the projectors $\prt{m}$}

The final form of the projection operators $\prt{m}$, defined in Eq.~(\ref{eq:prt}), is written in terms of a few coefficients $h_m$ and $e_n^{(m)}$. The former is explicitly given by
\begin{equation*}
h_0 = 6 (N + 3)(N + 4) (N + 6),
\end{equation*}
\begin{equation*}
h_1 = 3 (N + 1)(N + 3) (2 N + 3),
\end{equation*}
\begin{equation*}
h_2 = N (N + 1)(N + 2) (N + 4),
\end{equation*}
\begin{equation*}
h_3 = 2 N (N + 1)(N + 2),
\end{equation*}
\begin{equation*}
h_4 = N (N + 2)(N + 3) (N + 6),
\end{equation*}
and
\begin{equation*}
h_5 = N (N + 1)(N + 2) (N + 3) (2 N + 3),
\end{equation*}
and the latter is given by
\begin{equation*}
e_0^{(0)} = 0,
\end{equation*}
\begin{equation*}
e_1^{(0)} = - N (N + 1) (N + 2),
\end{equation*}
\begin{equation*}
e_2^{(0)} = 2 N^3 + N^2 - 6 N - 4,
\end{equation*}
\begin{equation*}
e_3^{(0)} = 2 N(5 N + 6),
\end{equation*}
\begin{equation*}
e_4^{(0)} = 4(4 N +3),
\end{equation*}
\begin{equation*}
e_5^{(0)} = 8,
\end{equation*}
\begin{equation*}
e_0^{(1)} = 0,
\end{equation*}
\begin{equation*}
e_1^{(1)} = 16 N (N^2 + 3 N + 2),
\end{equation*}
\begin{equation*}
e_2^{(1)} = - 8 (N^3 - 7 N^2 - 18 N - 8),
\end{equation*}
\begin{equation*}
e_3^{(1)} = - 8 (5 N^2 - 6 N - 12),
\end{equation*}
\begin{equation*}
e_4^{(1)} = - 64 N,
\end{equation*}
\begin{equation*}
e_5^{(1)} = - 32,
\end{equation*}
\begin{equation*}
e_0^{(2)} = 0,
\end{equation*}
\begin{equation*}
e_1^{(2)} = - 8 (N + 1) (N + 2),
\end{equation*}
\begin{equation*}
e_2^{(2)} = 4(5 N^2 + 9 N + 2),
\end{equation*}
\begin{equation*}
e_3^{(2)} = -4 (2 N^2 - 9 N - 12),
\end{equation*}
\begin{equation*}
e_4^{(2)} = - 8 (3 N - 1),
\end{equation*}
\begin{equation*}
e_5^{(2)} = - 16,
\end{equation*}
\begin{equation*}
e_0^{(3)} = 2 N (N^2 + 3 N + 2),
\end{equation*}
\begin{equation*}
e_1^{(3)} = - 5 N^3 - 5 N^2 + 10 N + 8,
\end{equation*}
\begin{equation*}
e_2^{(3)} = 2 N^3 - 19 N^2 - 30 N - 4,
\end{equation*}
\begin{equation*}
e_3^{(3)} = 2 (5 N^2 - 10 N - 12),
\end{equation*}
\begin{equation*}
e_4^{(3)} = 16 N - 4,
\end{equation*}
\begin{equation*}
e_5^{(3)} = 8,
\end{equation*}
\begin{equation*}
e_0^{(4)} = 0,
\end{equation*}
\begin{equation*}
e_1^{(4)} = 8 N (N + 1),
\end{equation*}
\begin{equation*}
e_2^{(4)} = - 4 (N - 1) (5 N + 4),
\end{equation*}
\begin{equation*}
e_3^{(4)} = 4 (2 N^2 - 13 N - 6),
\end{equation*}
\begin{equation*}
e_4^{(4)} = 24 (N - 1),
\end{equation*}
\begin{equation*}
e_5^{(4)} = 16,
\end{equation*}
\begin{equation*}
e_0^{(5)} = 0,
\end{equation*}
\begin{equation*}
e_1^{(5)} = - N (2 N + 4),
\end{equation*}
\begin{equation*}
e_2^{(5)} = 5 N^2 + 2 N - 8,
\end{equation*}
\begin{equation*}
e_3^{(5)} = -2 (N^2 - 8 N - 6),
\end{equation*}
\begin{equation*}
e_4^{(5)} = - 4 (2 N - 3),
\end{equation*}
\begin{equation*}
e_5^{(5)} = - 8.
\end{equation*}

\end{document}